\documentclass[10pt,aps,prb,superscriptaddress,twocolumn,floatfix]{revtex4-1}
\usepackage{graphicx}
\usepackage{dcolumn}
\usepackage{bm}
\usepackage{amsmath}
\usepackage{subfigure}
\usepackage{hyperref}
\usepackage{url}

\begin{document}

\preprint{APS/123-QED}

\title{Metastable solitonic states in the strained itinerant helimagnet FeGe}

\author{Victor Ukleev}
\email{victor.ukleev@psi.ch}
\affiliation{RIKEN Center for Emergent Matter Science (CEMS), Wako 351-0198, Japan}
\affiliation{Laboratory for Neutron Scattering and Imaging (LNS), Paul Scherrer Institute (PSI), CH-5232 Villigen, Switzerland}
\author{Yuichi Yamasaki}
\affiliation{RIKEN Center for Emergent Matter Science (CEMS), Wako 351-0198, Japan}
\affiliation{Research and Services Division of Materials Data and Integrated System (MaDIS), National Institute for Materials Science (NIMS), Tsukuba, 305-0047 Japan}
\affiliation{PRESTO, Japan Science and Technology Agency (JST), Kawaguchi 332-0012, Japan}
\author{Oleg Utesov}
\affiliation{Petersburg Nuclear Physics Institute NRC "Kurchatov Institute", Gatchina, Saint-Petersburg 188300, Russia}
\affiliation{St. Petersburg State University, 7/9 Universitetskaya nab., St. Petersburg 199034, Russia}
\affiliation{St. Petersburg Academic University - Nanotechnology Research and Education Centre of the Russian Academy of Sciences, 194021 St. Petersburg, Russia}
\author{Kiyou Shibata}
\altaffiliation[Current Affiliation: ]{Institute of Industrial Science, The University of Tokyo, Tokyo 153–8505, Japan}
\affiliation{RIKEN Center for Emergent Matter Science (CEMS), Wako 351-0198, Japan}
\author{Naoya Kanazawa}
\affiliation{Department of Applied Physics and Quantum-Phase Electronics Center (QPEC), University of Tokyo, Tokyo 113-8656, Japan}
\author{Nicolas Jaouen}
\affiliation{Synchrotron SOLEIL, Saint-Aubin, BP 48, 91192 Gif-sur-Yvette Cedex, France}
\author{Hironori Nakao}
\affiliation{Condensed Matter Research Center and Photon Factory, Institute of Materials Structure Science, High Energy Accelerator Research Organization, Tsukuba 305-0801, Japan}
\author{Yoshinori Tokura}
\affiliation{RIKEN Center for Emergent Matter Science (CEMS), Wako 351-0198, Japan}
\affiliation{Department of Applied Physics, University of Tokyo, Tokyo 113-8656, Japan}
\affiliation{Tokyo College, University of Tokyo, Tokyo 113-8656, Japan}
\author{Taka-hisa Arima}
\affiliation{RIKEN Center for Emergent Matter Science (CEMS), Wako 351-0198, Japan}
\affiliation{Department of Advanced Materials Science, University of Tokyo, Kashiwa 277-8561, Japan}

\begin{abstract}
The tensile strain is a promising tool for creation and manipulation of magnetic solitonic textures in the chiral helimagnets via tunable control of magnetic anisotropy and Dzyaloshinskii-Moriya interaction. Here, by using the in-situ resonant small-angle x-ray scattering we demonstrate that the skyrmion and chiral soliton lattices can be achieved as metastable states in FeGe lamella as distinct states under the tensile strain and magnetic fields in various orientations with respect to the deformation. The small-angle scattering data can be well accounted for in the frame of the analytical model for soliton lattice. By using the experimental results and analytical theory, unwinding of the metastable skyrmions in the perpendicular magnetic field as seen by small-angle scattering experiment was analyzed by the micromagnetic simulation.
\end{abstract}


\maketitle


\section{Introduction}

\begin{figure}
\includegraphics[width=7.5cm]{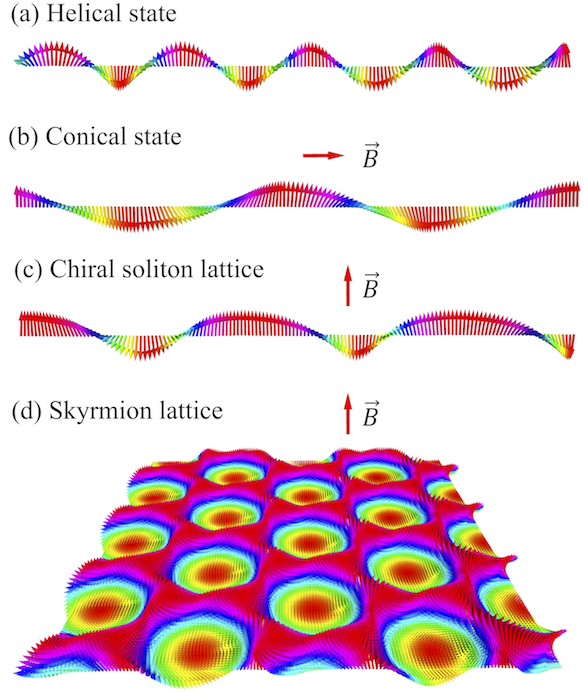}
\caption{Schematic illustration of spin configurations in helical state (a), conical state (b), chiral soliton lattice (c), and skyrmion lattice (d).}
\label{fig0}
\end{figure}

Antisymmetric Dzyaloshinskii-Moriya (DM) \cite{dzyaloshinsky1958thermodynamic,moriya1960anisotropic} and Heisenberg exchange interactions in cubic B20 chiral magnets (MnSi, FeGe, FeCoSi) result in a helical ground state \cite{bak1980theory} in which the magnetic moments in the neighboring atomic layers maintain a fixed propagating angle thus forming a spin helix with the fixed chirality (Fig. \ref{fig0}a). The helical wavelength $\lambda=4\pi A_{ex}/D$ is determined by the ratio of the exchange stiffness $A_{ex}$ and the Dzyaloshinskii constant $D$. At zero field, the propagation vector of the helix $\mathbf{q}$ is imposed by a cubic anisotropy and, typically, encompassed along $\langle111\rangle$ (MnSi, FeGe) or $\langle100\rangle$-equivalent (MnGe, Cu$_2$OSeO$_3$) axes \cite{bak1980theory}. Application of a magnetic field parallel to the $\mathbf{q}$-vector results in canting of the magnetic moments towards the field direction, and the helical texture transforms to a conical state (Fig. \ref{fig0}b). Further increment of magnetic field leads to the transition from the conical to induced ferromagnetic state. In cubic B20 helimagnets, the orientation of the helical $q$-vector is determined by the relatively weak cubic anisotropy and can be rotated by a small magnetic field. In the case of strong uniaxial anisotropy that can be caused, for example, by the compressive or tensile strain, a moderate magnetic field applied perpendicular to the helical propagation axis can deform the proper screw magnetic modulation into a chiral soliton lattice (CSL) \cite{dzyaloshinskii1965theory,izyumov1984modulated} (Fig. \ref{fig0}c). Typically, a CSL appears in the uniaxial chiral magnets due to the interplay between easy-axis type anisotropy, DMI interaction and external magnetic field. Recently, the interest in magnetic CSLs has been lifted by the theoretical predictions of driving magnetic kinks by electric current and crossed magnetic fields \cite{bostrem2008theory,koumpouras2016spin}, and non-trivial dynamics of the solitons in the GHz frequency range \cite{togawa2016symmetry,goncalves2017collective}. Following the theoretical conceptions, several experimental works unambiguously demonstrated formation and manipulation of the CSL in prototypical uniaxial chiral helimagnet CrNb$_3$S$_6$ \cite{togawa2012chiral,togawa2013interlayer,yonemura2017magnetic,tabata2020observation}. 

Another peculiar magnetic structure that appears in the chiral helimagnets is a skyrmion crystal (SkX), a hexagonally ordered array of topologically protected chiral magnetic vortices \cite{muhlbauer2009skyrmion,yu2010real} (Fig. \ref{fig0}d). Recently, Shibata et al. have shown that the SkX and individual skyrmions can be deformed by the tensile strain induced by the lattice mismatch between FeGe lamella and silicon substrate and this phenomena was explained by anisotropic change of DM interaction \cite{shibata2015large}. Furthermore, transformation of the magnetic ground state of Cu$_2$OSeO$_3$ from proper-screw structure to CSL upon application of the tensile strain, also accompanied by the deformation of the SkX has been demonstrated by Okamura et al. \cite{okamura2017emergence}. In the latter case, the transformation of the ground state was explained by change of magnetic anisotropy induced by the strain. Recently, a large enhancement of the SkX stability in the chemically strained bulk B20 helimagnet MnSi has been demonstrated \cite{sukhanov2019giant}.
Tensile strain induced by the lattice mismatch between the B20 crystal epilayers and the substrate has been also imposed in the thin film samples of MnSi and FeGe \cite{wilson2012extended,huang2012extended,li2013robust}. The strain effect may either extend \cite{wilson2012extended,huang2012extended} or destabilize the skyrmion phase rather favoring formation of the CSL \cite{porter2015manipulation,kanazawa2016direct,zhang2017room}. Therefore, these induced uniaxial anisotropy effects on chiral helimagnets provide a fertile ground for investigation of the interplay between magnetic interactions, leading to the rich variety of ground and metastable magnetic states. 

Recently, there have been a number of reports on the metastable SkX formation by quenching the sample under applied magnetic field \cite{oike2016interplay,karube2017skyrmion,yu2018aggregation,nakajima2018uniaxial,qian2018new,white2018electric,chacon2018observation,wilson2020stability}. By utilizing this technique, a robust skyrmion state can be obtained in a thin plate of FeGe at low temperatures even at zero magnetic field, providing various opportunities for investigations of SkX stability under oblique, perpendicular and negative magnetic fields. In this paper we report on a resonant small-angle x-ray scattering (RSXS) study of the polymorphic magnetic states in a thin strained FeGe lamella. Experiment reveals that the combination of the in-plane magnetic field and the tensile strain stabilizes CSL with the $q$-vector perpendicular to the strain direction. Furthermore we demonstrate that depending on the magnetic field direction and cooling protocol, the hysteretic transformations and the coexistence of metastable modulated states, such as SkX and CSL are observed at low temperatures.

\section{Experimental}

\begin{figure}
\includegraphics[width=8.5cm]{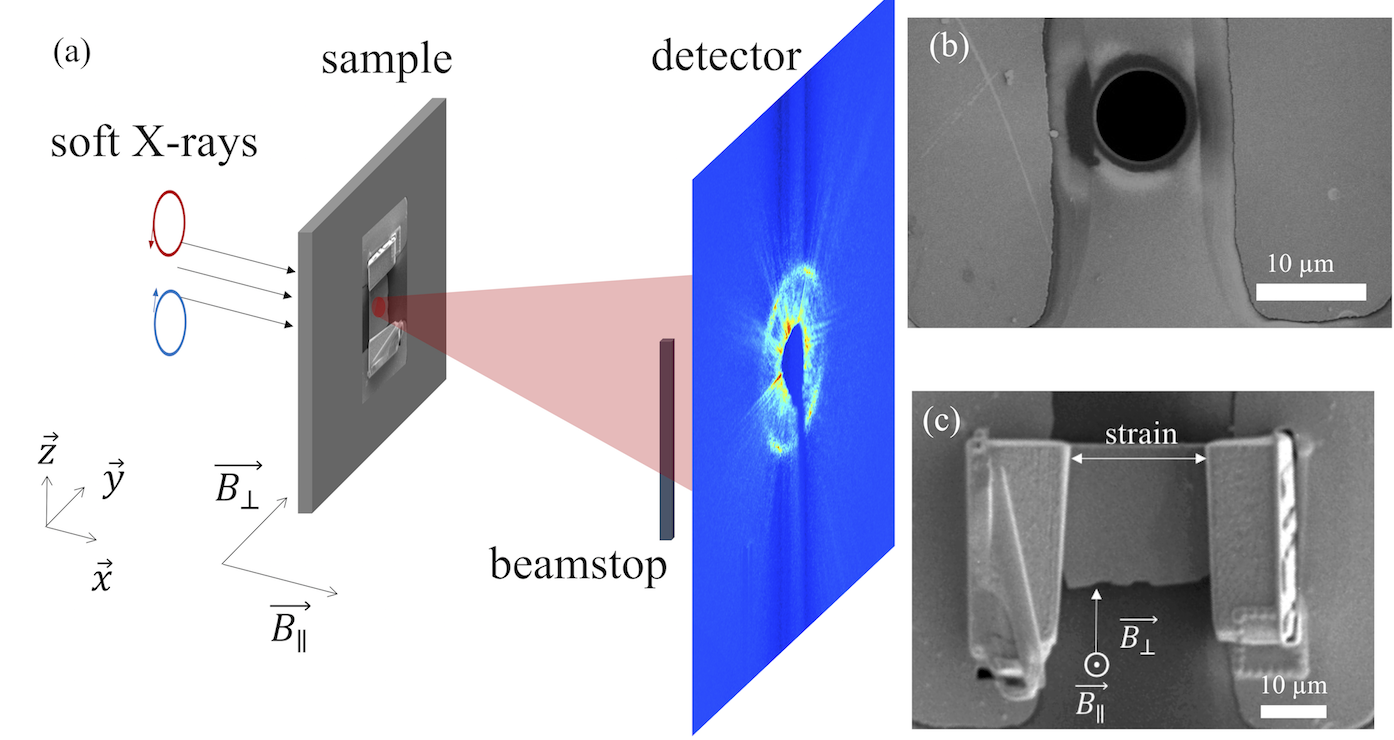}
\caption{(a) Schematic of the RSXS experiment. SEM image of the (b) sample aperture and (c) thin plate of FeGe fixed onto the membrane. Note that in the actual experiment the detector plane and the beamstop were rotated by $-45^\circ$ to $z$ axis in the $(yz)$ plane and magnetic field $B_{\bot}$ was applied by rotating the sample plane around $z$ axis.}
\label{fig1}
\end{figure}

The RSXS experiments were carried out at the soft x-ray beamlines BL-16A, KEK Photon Factory (Tsukuba, Japan) and SEXTANTS, SOLEIL (Gif-sur-Yvette, France). A sketch of the experimental geometry at BL-16A and SEM images of the sample are given in Fig. \ref{fig1}. The experiments were performed with the circularly polarized soft x-rays of a photon energy $E = 707$\,eV corresponding to the $L_3$ absorption edge of Fe. The small-angle scattering setup at BL-16A was equipped with a high-vacuum chamber with a background pressure of $10^{-8}$\,Torr \cite{yamasaki2015dynamical}. The intensities of scattered beams were collected by an in-vacuum charge coupled device (CCD) detector of $512\times512$\,pixels (Princeton Instruments, Trenton, New Jersey, USA) protected from the direct beam by a tungsten beamstop. A pair of Helmholtz coils provided a static magnetic field $B_{\parallel}$ in the direction parallel to the incident beam. A magnetic field $B_{\bot}$ parallel to the sample plane was applied by rotating the sample by $90^\circ$ (i.e. around $z$ axis in Fig.\ref{fig1}a). Therefore the RSXS patterns in the present study were measured at $B_{\bot}=0$ after ex-situ application of $B_{\bot}$. The magnitude of a magnetic field provided by the electromagnet was in the range from 0 to 400\,mT. The sample temperature was controlled by a He-flow-type refrigerator.
The RESOXS setup at SEXTANTS was equipped with a quadruple magnet that allows the in-situ observation of the scattering patterns for both in-plane and out-of-plane magnetic fields geometries in the field range from 0 to 150\,mT \cite{jaouen2004apparatus}. The data measured using RESOXS setup is given in the Appendix A.

The single-crystal sample of FeGe was grown by chemical vapor transport method \cite{kanazawa2015magnetic}. The lamella with the thickness of 150 nm was prepared by focused ion beam (FIB) milling and fixed to a gold-coated silicon nitride membrane behind the aperture with a diameter of 8\,$\mu$m (Fig. \ref{fig1}b). Tensile strain occurred due to the tungsten deposition at the both sides of the thin plate and the difference of the thermal expansion coefficients between FeGe and Si$_3$N$_4$ membrane \cite{yamasaki2015dynamical,okamura2017emergence}. The strain direction is shown in Fig. \ref{fig1}c.

\begin{figure*}
\includegraphics[width=17cm]{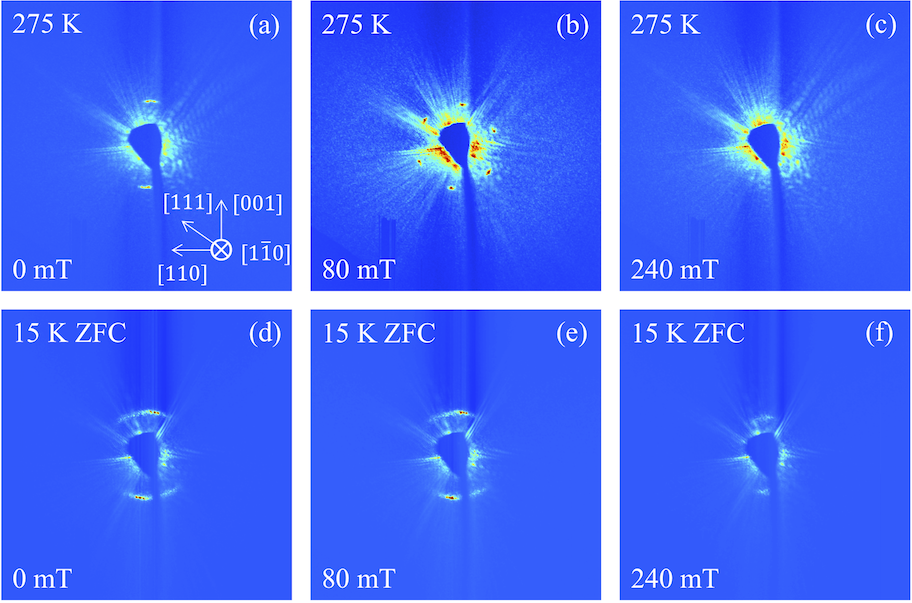}
\caption{Typical RSXS patterns for strained FeGe sample. (a -- c) At $T=275$\,K the helical state shown in (a) transforms to (b) SkX at 50--140\,mT and (c) conical / induced ferromagnetic state at $150$\,mT. (d --f) After zero-field cooling down to $T=15$\,K the helical state shown in (d) gradually transforms to the conical / induced ferromagnetic state with increasing magnetic field.}
\label{fig2}
\end{figure*}

\section{Results}

At 275\,K the typical RSXS patterns arising from the helical, SkX and conical / field-polarized phases of FeGe \cite{yamasaki2015dynamical,ukleev2018coherent,burn2019helical,burn2020field} were observed (Figs. \ref{fig2}a--c). The orientation of the helical $q$-vector was pinned along the [001], while the orientation of the SkX was not fixed and the skyrmion lattice stochastically transformed between the single-domain and the multi-domain states, whereas the relative rotation angle between domains was approximately $15^\circ$. This rotation indicates that in the present sample the magnitude of the tensile strain at $T=275$\,K was not sufficient to lock the orientation of the SkX as it was observed in Refs.\cite{shibata2015large,yamasaki2015dynamical,okamura2017emergence,okamura2017directional}. Dependence of magnetic scattering on magnetic field $B_{\parallel}$ was measured at the lowest temperature $T=15$\,K after zero-field cooling (ZFC) procedure (Fig. \ref{fig2}d--f). ZFC led to the transition from the single-domain helical to the conical or field-polarized magnetic structure as the field increased from 0 to 150\,mT or above. No higher-order harmonics of the magnetic scattering were observed indicating direct helical-to-conical transition without appearance of the intermediate CSL or SkX states. By ramping of magnetic field $B_{\parallel}$ to the maximal value of 400\,mT and going back to zero field, two helical domains with the propagation vectors encompassed along the [111]-equivalent directions (Fig. \ref{fig7} in Appendix B) appear. 
Intensity of the Bragg peak arising from the helical domain with the $q$-vector parallel to the strain is approximately five times stronger than the one with the perpendicular propagation direction. This indicates the enhanced stability of the magnetic helices with the $q$-vector along the tensile strain. Therefore we assume the easy-plane type anisotropy induced by the deformation with the anisotropy axis parallel to the strain. 

\begin{figure*}
\includegraphics[width=17cm]{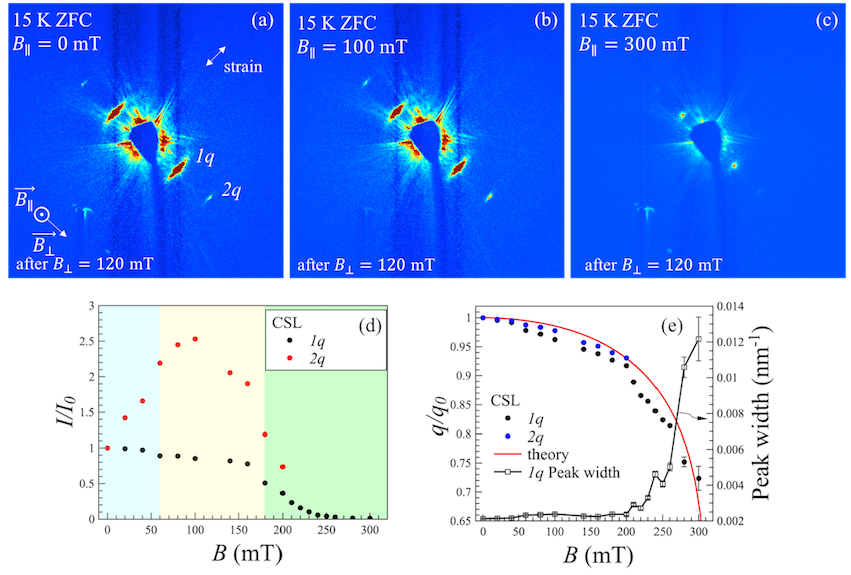}
\caption{RSXS patterns of strained FeGe lamella measured at $T=15$\,K after application and removal of $B_{\bot}=120$\,mT at (a) $B_{\parallel}=0$\,mT, (b) $B_{\parallel}=100$\,mT, (c) $B_{\parallel}=300$\,mT. First- and second-order Bragg peaks arising from CSL are indicated as $1q$ and $2q$, respectively. (d) Dependence of the relative intensity of the $1q$ and $2q$ peaks on applied magnetic field $B_{\parallel}$. (e) Dependence of the relative $q$-vector magnitude and fwhm of the $1q$ and $2q$ peaks on applied magnetic field $B_{\parallel}$.}
\label{fig3}
\end{figure*}

\begin{figure*}
\includegraphics[width=17cm]{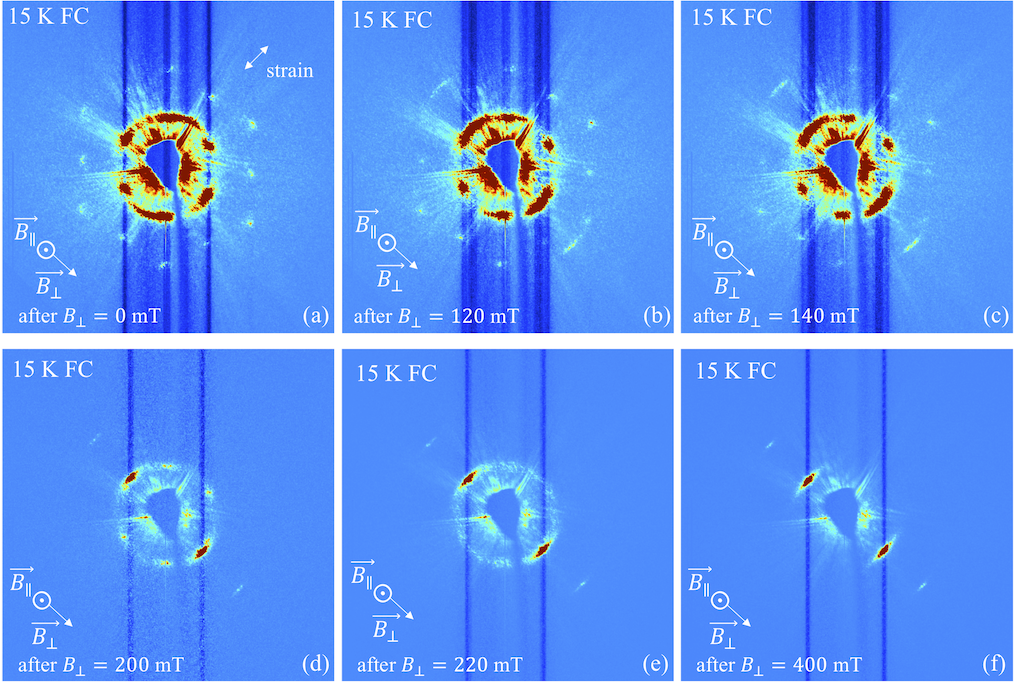}
\caption{RSXS patterns for strained FeGe lamella measured at $T=15$\,K and zero field condition after field cooling in $B_{\parallel}=120$\,mT and applied in-plane magnetic field  (a) $B_{\bot}=0$\,mT, (b) $B_{\bot}=120$\,mT, (c) $B_{\bot}=140$\,mT, (d) $B_{\bot}=200$\,mT, (e) $B_{\bot}=220$\,mT, (f) $B_{\bot}=400$\,mT.}
\label{fig4}
\end{figure*}

Surprisingly, application and consequent withdrawal of the magnetic field $B_{\bot}=120$\,mT perpendicular to the tensile strain in the sample plane leads to appearance of second-order Bragg reflections (Fig. \ref{fig3}a), which unambiguously indicates an anharmonicity of the magnetic modulation. Similar higher-order scattering peaks have been previously observed in other chiral magnets \cite{grigoriev2006magnetic,okamura2017emergence,nakajima2018uniaxial} and centrosymmetric systems with long-periodic magnetic texture \cite{durr1999chiral,hellwig2003x,desautels2019realization} and can be applied as indicators of a spiral distortion. Therefore, the formation of a distorted helical texture, or CSL was assisted by the magnetic field applied perpendicular to the strain in the thin plate plane, which has not been reported previously for FeGe or Cu$_2$OSeO$_3$. Observation of the CSL at zero field indicates the metastable character of this state. By applying the field $B_{\parallel}$ normal to the sample surface, the evolution of the amplitudes and positions of the first-order ($1q$) and second-order ($2q$) magnetic Bragg peaks was recorded (Fig. \ref{fig3}). Interestingly, the CSL was not destroyed by the magnetic field, indicating rotation of the non-zero in-plane net magnetization component normal to the plane. We note, that in case of the helical re-orientation towards the applied $B_{\parallel}$ field the diffraction intensity would gradually decrease without manifestation of the higher harmonics. Therefore, instead of the usual helical-to-conical transition we assume that the conical state competes with the metastable CSL when the field $B_{\parallel}$ sequentially applied after $B_{\bot}$. When the field is increased above 200 mT the CSL transforms to the conical state with the $q$-vector oriented parallel to the field direction. This field-induced transition from the in-plane helical to the out-of-plane conical state is mediated by the in-plane CSL, which is different from, for example, uniaxial helimagnet CrNb$_3$S$_6$ where conical phase is suppressed and helical-to-field-polarized transition takes place via CSL phase \cite{togawa2012chiral}. Due to the gradual variation of the intensity, magnitude of the $q$-vector and the full width at half maximum (fwhm) of the second-order Bragg peaks (Figs. \ref{fig3}d,e), we suggest the phase co-existence between the CSL and conical domains between 100 and 250\,mT. The magnetic field dependence of $q$-vector in this process consists with the chiral sine-Gordon model \cite{dzyaloshinskii1965theory,izyumov1984modulated} (Fig. \ref{fig3}e) for the lower field range ($0<B<100$\,mT). Notably, an field in-plane applied parallel to the strain axis does not induce the CSL, but leads to the gradual rotation of the propagation vector towards the field (see the Appendix A). More accurate description of the CSL flop above 100\,mT is given in the Discussion section.

\begin{figure}
\includegraphics[width=8.5cm]{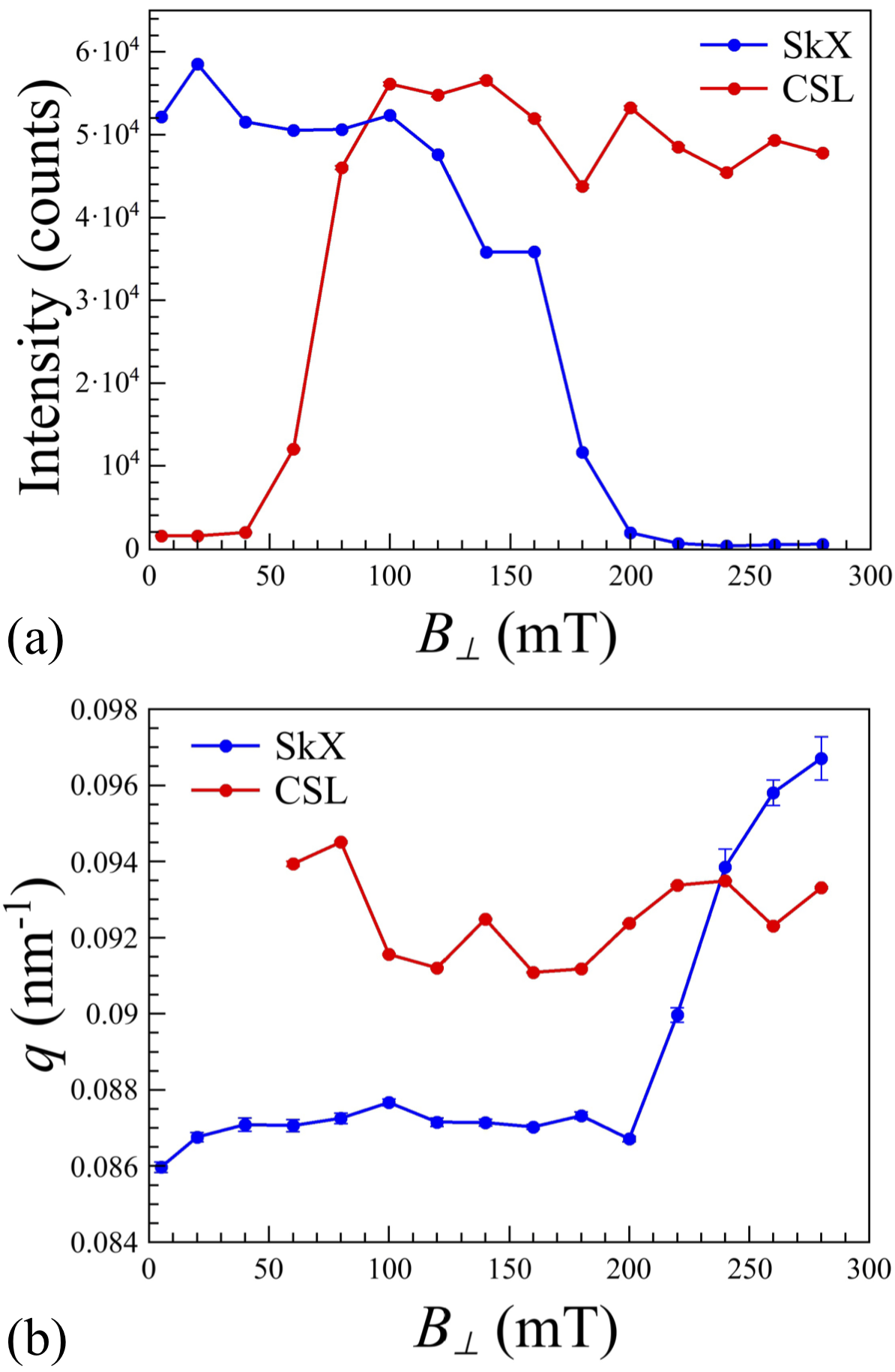}
\caption{In-plane magnetic field dependence of (a) the intensity and (b) $q$-vector magnitude of the Bragg peaks corresponding to the skyrmion lattice (SkX) and chiral soliton lattice (CSL).}
\label{fig5}
\end{figure}

Next we explore the field-tuning between a metastable SkX state and CSL states. Field cooling (FC) from 275 K to 15 K in the applied magnetic field $B_{\parallel}=120$\,mT with the cooling rate of 0.1-0.4\,K/s led to appearance of the metastable SkX which remained even in zero field (Fig. \ref{fig4}a). The magnetic contribution to RSXS dramatically increased at the low-temperature conditions, providing the access to the higher-order Bragg peaks from SkX. Stability of the supercooled SkX over the perpendicular magnetic field has been studied with the protocol described in the Experimental section. Evolution of the magnetic scattering intensity with $B_{\bot }$ is shown in Fig. \ref{fig4}. In zero field the SkX co-exists with the helical domain (Fig. \ref{fig4}a). After application of the in-plane magnetic field in a range from 20\,mT to 200\,mT, the magnetic texture, firstly, evolves into the co-existing SkX and CSL oriented along $B_{\bot}$ (Figs. \ref{fig4}(b--d)), as it is suggested by six-fold pattern arising from SkX and $1q$ and $2q$ Bragg peaks from CSL. Domination of the metastable CSL over the supercooled SkX proceeds when increasing the in-plane field, and also lowers the coherence of the SkX, as it is indicated by vanishing of the higher-order Bragg peaks of SkX [Fig. \ref{fig4}(d)]. After application of a field $B_{\bot}=220$\,mT a ring-like pattern coexistent with the diffraction spots from the CSL was observed. This pattern is attributed to the disordered skyrmion lattice or helical domains. Dependencies of the integrated intensity and magnitude of the $q$-vector of the first-order Bragg peaks of SkX and CSL are summarized in Figs. \ref{fig5}a,b. The low and intermediate-field regions ($0<B_{\bot}<200$\,mT) correspond to the co-existing SkX and helical/CSL magnetic textures. Next, the intensity dependence in the higher-field region ($200<B_{\bot}<300$\,mT) demonstrate mixed CSL and disordered skyrmion phases, although the scattering intensity from the SkX is vanishing compared to the single-$q$ modulation (Fig. \ref{fig5}a). Surprisingly, the $q$-vector length of the skyrmion texture tends to increase with the increment of magnetic field (Fig. \ref{fig5}b), which is opposite to the tendency for the SkX, isolated skyrmions or in-plane skyrmion tubes in the conical/ferromagnetic background \cite{bogdanov1994thermodynamically,mcgrouther2016internal,birch2020real}. By contrast, application of the in-plane field does not influence the magnitude of the CSL wavevector (Fig. \ref{fig5}b). Application of the in-plane field above 300\,mT completely destroys the skyrmion crystal and only the CSL remains. 

\section{Discussions}

Recently, Lorentz transmission electron microscopy (LTEM) experiments showing a thickness-dependent stability of a helicoidal, conical and skyrmion states in FeGe lamella with thickness gradient in the presence of magnetic field have been discussed in Ref. \cite{leonov2016chiral} in context of chiral surface twists. In the present Discussion section we focus on a theoretical model which is sufficient to describe the experimental RSXS data obtained for zero-field cooling case, i.e., considering only helical, conical and CSL phases. To discuss the experimental data obtained for the metastable SkX, a micromagnetic simulation has been performed.

\subsection{Chiral soliton lattice}

Chiral soliton lattices are usually attributed to uniaxial helimagnets, where spiral can propagate only along single direction and spins are collinear within the planes that are perpendicular to the axis. In this subsection we recall the classical solutions for uniaxial helimagnets in external in-plane magnetic field or with additional in-plane easy-axis anisotropy (provided, e.g., by tensile stress), which lead to the proper screw distortion.

Here we consider a simple model of uniaxial helimagnet on tetragonal lattice, assuming an isotropic ferromagnetic exchange which is the same in all directions and Dzyaloshinskii-Moriya interaction along $z$-axis between nearest neighbors in neighboring $xy$-planes. Additionally, we introduce to the model an in-plane magnetic field and an in-plane easy-axis anisotropy. The corresponding system Hamiltonian reads
\begin{equation}
 \begin{aligned}
  \label{H0start}
   \mathcal{H}=-J \sum_{\langle i j \rangle} {\bf S}_{i} \cdot {\bf S}_{j}- \sum_{[ij]} {\bf D}\cdot \left[ {\bf S}_{i}\times {\bf S}_{j} \right] - \\ - \sum_{i} g \mu_B \mathbf{H} \cdot \mathbf{S}_{i} - A \sum_i  \mathbf{S}^2_{x,i},
 \end{aligned}
\end{equation}
henceforth we consider magnetic field in energy units, $h=g \mu_B H$. In the absence of anisotropy and external field the ground state of this system is the spin helix propagating along $z$-axis with spins rotating in $xy$-plane and modulation vector $\mathbf{q} = ( 0, 0, D/J)$. When $q \ll 1$ one can rewrite Hamiltonian~\eqref{H0start} in continuous form using variable $\varphi(z)$ which corresponds to spin polar angle in $xy$-plane.

For the system in external effective in-plane magnetic field $\alpha=h/SJ$ the solution is given by the following expressions \cite{izyumov1984modulated}:
\begin{eqnarray}
  \label{Phi1}
  \varphi(z)=2 \, \text{am}\!\left( \sqrt{\frac{\alpha}{m}}z, m  \right), \\
  \label{Lkink3}
   L_{kink} = \frac{8 K(m) E(m)}{\pi q}, \\
   \label{Lkink2}
   \frac{4 E(m)}{\pi q} = \sqrt{\frac{m}{\alpha}}.
\end{eqnarray}
Here $m$ is the parameter, which should be found from the last equation, $K(m)$ and $E(m)$ are the complete elliptic integrals of the first and the second kinds, respectively, and $\text{am}$ is Jacobi amplitude function. These equations yield the periodicity of magnetic modulation $L_{kink}=2\pi/q$ for the zero external field ($\alpha=0$, $m=0$). The phase transition to the fully polarized state takes place at $m=1$, for which $\alpha_c=(\pi q/4)^2$. At small magnetic fields $\alpha \ll \alpha_c$ we find $m= 4 \alpha / q^2 (1- 2\alpha/q^2)$ and corresponding expression for the system energy per one spin is given by
\begin{equation}\label{En1}
  \frac{\varepsilon}{N} = - 3 S^2 J - \frac{S^2 D^2}{2 J} - \frac{J h^2}{4 D^2}.
\end{equation}

In case of non-zero in-plane easy-axis anisotropy ($A\neq 0$) the model can be directly mapped onto the previous case with $\tilde{q}=2 q$ and $\tilde{\alpha}= 2A /J$. Considering anisotropy as a small perturbation ($A \ll D^2/J$) we can derive the energy of the corresponding magnetic structure per one spin in the following form:
\begin{equation*}\label{En2}
   \frac{\varepsilon}{N} = - 3 S^2 J  - \frac{S^2 D^2}{2 J} - \frac{S^2 A}{2} - \frac{S^2 J A^2}{4 D^2}.
\end{equation*}

Now we consider an external magnetic field along $z$-direction. Assuming that all spins are canted to the same angle $\theta$ towards the field direction, after the minimization of energy per one spin over $\theta$, we obtain
\begin{equation}
\label{En4}
  \frac{\varepsilon}{N}= - 3 S^2 J -\frac{S^2}{2} \left(\frac{ D^2}{J} + A + \frac{ J A^2}{2 D^2} \right) - \frac{h^2}{2 \left[ \frac{D^2}{J} + A + \frac {J A^2}{2 D^2} \right]}
\end{equation}

In this form this model can be applied to analyze the cubic helimagnets with tensile strain along one of the directions.

\begin{figure*}[t]
\includegraphics[width=17cm]{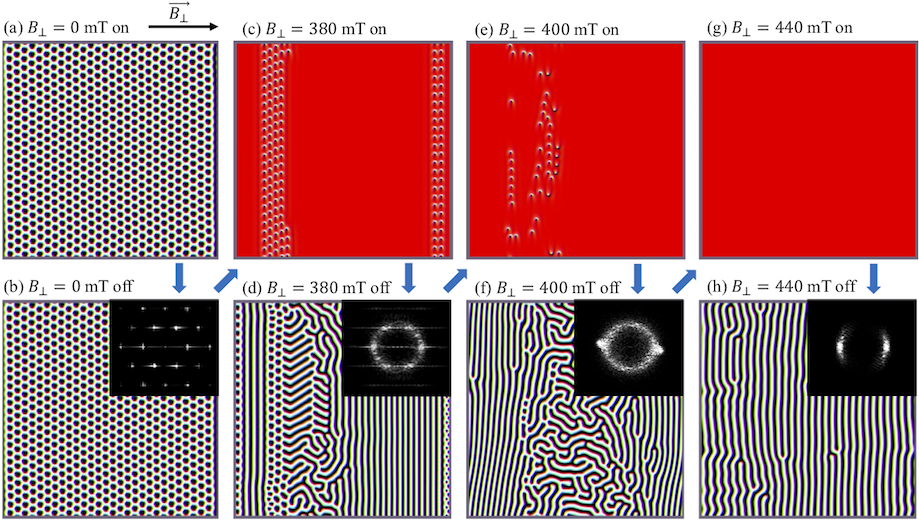}
\caption{Micromagnetic simulation of a FeGe lamella with uniaxial anisotropy after application and removal of in-plane magnetic fields (a,b) $B_{\bot}=0$\,mT, (c,d) $B_{\bot}=380$\,mT, (e,f) $B_{\bot}=400$\,mT, (g,h) $B_{\bot}=440$\,mT. Sequence of the field application protocol is denoted by blue arrows. In-plane magnetic field was consequently increased with 20\,mT step (``$B_{\bot}$ on'' state) and switched to zero (``$B_{\bot}$ off'' state). Insets show fast Fourier transform (FFT) images of the zero-field magnetization states after the corresponding in-plane field training for qualitative comparison with experimental RSXS patterns (Fig. \ref{fig4}). }
\label{fig6}
\end{figure*}

\subsection{Spiral plane flop}

In this subsection we study a simple model related to strained B20 helimagnets. It allows to describe spiral plane flop observed experimentally (Fig. \ref{fig3}). In this case we consider only nearest-neighbor exchange interaction and DMI with vectors along the corresponding bonds. The corresponding model Hamiltonian reads
\begin{eqnarray}
 \nonumber
  \mathcal{H} &=& \mathcal{H}_{ex}+\mathcal{H}_{dm}+\mathcal{H}_{an} + \mathcal{H}_{Z}, \\ \nonumber
  \mathcal{H}_{ex} &=& -\sum_{\langle i j \rangle} J_{ij} \mathbf{S}_i \cdot \mathbf{S}_i, \\ \label{HMnSi}
  \mathcal{H}_{dm}&=& -  \sum_{\langle i j \rangle} {\bf D}_{ij} \cdot [\mathbf{S}_i \times \mathbf{S}_j], \\ \nonumber
  \mathcal{H}_{an} &=&- A \sum_i \left( S^2_{x,i} + S^2_{y,i}\right), \\  \nonumber
  \mathcal{H}_Z &=& - \sum_i \mathbf{h} \cdot \mathbf{S}_{i}.
\end{eqnarray}
Here we consider the B20 cubic helimagnet with tensile strain along $z$-axis. This yields $xy$ easy-plane, and difference between exchange and DMI for in-plane bonds and along $z$-axis, namely $J, D$ and $J^\prime, D^\prime$.

If we apply a magnetic field along $x$-axis, two competing spin textures can arise: CSL of $\mathbf{q}=(0,0,q)$ with spins in $xy$-plane and bunched cone  of $\mathbf{q}=(q,0,0)$ with spin rotating in $yz$-plane. The former is energetically favorable at small magnetic fields, while the latter becomes stable as the field increases due to spin canting towards magnetic field direction. Analytically this stems from competition between two energies:
\begin{eqnarray}
  \label{enxy}
  \frac{\varepsilon_{xy}}{N} &=& - 2 S^2 J -S^2 J^\prime- \frac{S^2 D^{\prime 2}}{2 J^\prime} - S^2 A - \frac{J^\prime h^2}{4 D^{\prime 2}}, \\ \label{enyz}
  \frac{\varepsilon_{yz}}{N} &=& - 2 S^2 J - S^2 J^\prime- \frac{S^2 D^2}{2 J} - \frac{S^2 A}{2} - \frac{ J S^2 A^2}{4 D^2} -  \nonumber \\
& &  -\frac{h^2}{2 \left[ A + D^2/J +J A^2/(2 D^2) \right]}.
\end{eqnarray}
For clarity we consider specific situation. Let $D^2/J = D^{\prime 2}/ J^\prime$, and anisotropy constant $A \ll D^2/J$ is small. It is obvious from the Eqs.~\eqref{enxy} and~\eqref{enyz} that the energy of the bunched cone texture decreases in magnetic field twice faster than the CSL one, and at a sufficiently large external field a spiral plane flop takes place. The transition point is estimated by the equation
\begin{equation}\label{Hflop1}
  \frac{AS^2}{2}=\frac{J h_{flop}^2}{4 D^2}.
\end{equation}
This gives
\begin{equation}\label{Hflop2}
  h_{flop}=S\sqrt{\frac{2 A D^2}{J}}.
\end{equation}
We can easily confirm that $h_{flop} \ll h_c \propto S D^2/J$, which justifies the small field approach for CSL energy. By using the measured ratio $h_{flop}/h_{c}\approx 1/3$ this equation leads to the estimation of the induced easy-plane anisotropy constant $A\approx 1.6\cdot10^4$\,J/m$^3$.

\subsection{Micromagnetic simulation of the skyrmion lattice in the in-plane magnetic field}

By assuming the easy-plane anisotropy axis along the deformation direction and magnitude $A=1.6\cdot10^4$\,J/m$^3$ estimated from the analytical theory and experimental results, we performed the simulation of the skyrmion lattice in the strained FeGe lamella by using mumax$^3$ package \cite{vansteenkiste2014design}. In simulation we used two-dimensional thin plate $2048\times2048$\,nm$^2$ and thickness of 100\,nm with periodic boundary conditions along the in-plane directions. The parameters used for the simulation were obtained experimentally by microwave spin-wave spectroscopy in Ref. \cite{takagi2017spin}. Metastable SkX was introduced as the initial zero-field state (Figs. \ref{fig6}a,b). The in-plane magnetic field deforms the metastable skyrmions resulting in elongation towards the field direction. By increasing the magnitude of the magnetic field aligned chains of skyrmions show up in the ferromagnetic background at $B_{\bot}=380$--$400$\,mT (Fig. \ref{fig6}c,e). Interestingly, when the field is released, both elongated and compressed skyrmion vortices of smaller diameter appear at the boundaries between the helical and skyrmion chain domains (Fig. \ref{fig6}d). Finally, by applying field of $B_{\bot}=440$\,mT the sample undergoes a transition to the homogeneously magnetized state (Fig. \ref{fig6}g). The procedure of consequent application/removal of the in-plane magnetic field reproduces the experimental conditions described in the previous sections: the in-plane magnetic field $B_{\bot}$ was consequently applied to initial zero-field SkX state with increasing 20\,mT step and switched off. By using this protocol, the micromagnetic simulation delivers skyrmion clusters and individual skyrmions co-existing with the helical domains with the $q$-vector oriented along the in-plane field direction (Figs. \ref{fig6}d,f). For comparison with experimental RSXS data the corresponding fast Fourier transform (FFT) patterns were calculated for the out-of-plane magnetization component of the resultant zero-field states (insets in Figs. \ref{fig6}b,d,h,f).

We suppose that the ring-like RSXS patterns observed in the experiment (Figs. \ref{fig4}c,f) may correspond to the mixed skyrmion clusters and multidomain helical (maze-like pattern in Figs. \ref{fig6}d,h,f) states delivered by the micromagnetic simulation. In our case the mixed skyrmion clusters and disordered helical pitches co-exist with the metastable CSL discussed in the previous sections. Although the simulation does not allow to reproduce the specific incremental $q$-dependence on magnetic field, the attractive nature of the skyrmion clusters in the helical background has been already observed by LTEM \cite{muller2017magnetic}. The skyrmion condensation is assisted by the non-axisymmetric skyrmion shapes induced by the in-plane magnetic field \cite{loudon2017direct}. The specific compressed skyrmions formed at the domain boundary between the skyrmion cluster and helical domain hint the possible scenario of the $q$-vector increase observed in the RSXS experiment (Fig. \ref{fig5}b). Contributions of more exotic three-dimensional spin textures, such as stacked spirals \cite{rybakov2016new}, chiral bobbers \cite{rybakov2015new,zheng2018experimental,ahmed2018chiral}, quasi-monopoles \cite{muller2020coupled}, and in-plane skyrmion tubes \cite{leonov2018toggle,vlasov2020skyrmion,birch2020real} to the scattering patterns is another possible explanation of this result. Further real-space investigations of the possible three-dimensional states using cryogenic coherent \cite{ukleev2018coherent,ukleev2019element} or focused \cite{birch2020real} soft x-ray methods will help to clarify this question.

\begin{figure*}
\includegraphics[width=17cm]{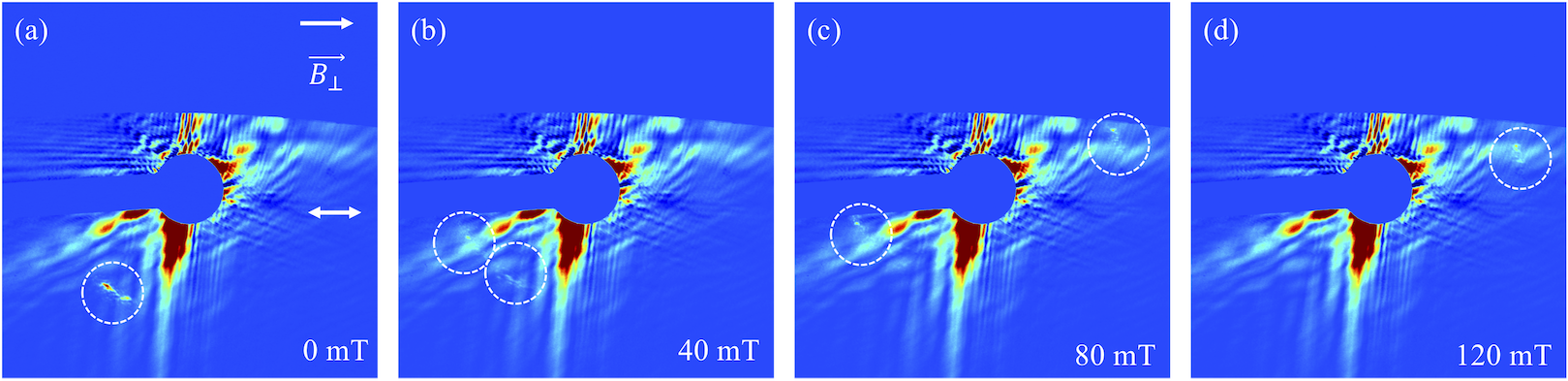}
\caption{RSXS patterns for strained FeGe lamella measured at $T=32$\,K after zero field cooling and in-plane field (a) $B_{\bot}=0$\,mT, (b) $B_{\bot}=40$\,mT, (c) $B_{\bot}=80$\,mT, (d) $B_{\bot}=120$\,mT applied parallel to the tensile strain direction. Magnetic Bragg peaks are highlighted by the dashed circles.}
\label{fig8}
\end{figure*}

\begin{figure}
\includegraphics[width=7.5cm]{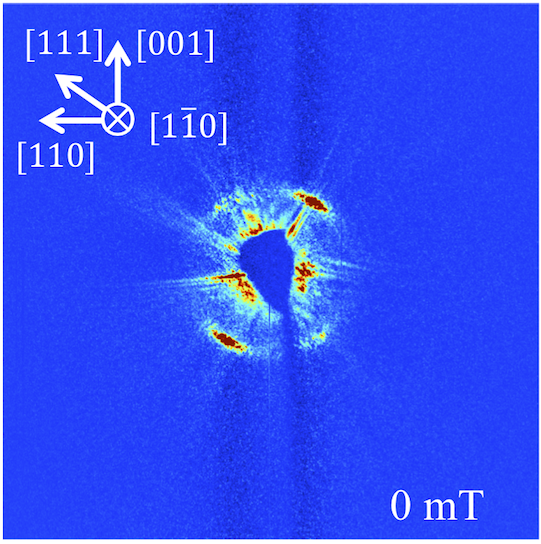}
\caption{RSXS pattern for strained FeGe lamella measured at $T=15$\,K and zero field condition after zero field cooling and out-of-plane field training at $B_{\parallel}=400$\,mT.}
\label{fig7}
\end{figure}

\section{Conclusion}

We have investigated the low-temperature magnetic states in a strained lamella of FeGe by means of resonant small-angle soft x-ray scattering. If the magnetic field is applied in the plane of lamella and perpendicular to the tensile strain direction at 15 K the chiral soliton lattice shows up as a metastable state when the field is removed. The skyrmion lattice is stabilized by the field cooling and survives even at zero field at low temperatures. Further application of the in-plane magnetic field results in the co-existence of the metastable skyrmion and chiral soliton lattices. Unwinding of the skyrmion crystal to the helicoidal state passing through the intermediate phase consisting of the isolated skyrmion clusters and disordered helices. This polymorphism of the low-temperature zero-field magnetic states results from the complex landscape of magnetic interactions, including exchange, anisotropy and Dzyaloshinskii-Moriya interactions. Finally, our measurements have shown non-trivial interaction between the co-existing metastable skyrmion and chiral soliton lattices, namely unexpected attractive behavior in skyrmion clusters. The revealed effect of the tensile strain can be potentially employed to control skyrmion and chiral soliton lattices in helimagnetic materials.
By modifying the anisotropy and Dzyaloshinskii-Moriya interaction in the cubic chiral magnets one can stabilize the rich variety of magnetic solitonic textures. It is also possible that the strain tuning can help to stabilize skyrmion lattice in the chiral helimagnets with lower crystal symmetry.

\section*{Acknowledgements}

Soft x-ray scattering experiments were performed at KEK Photon Factory as a part of the Projects No.: 2015S2-007 and 2018S2-006, and at synchrotron Sol\'eil as a part of the Proposal No. 20181292. This research was supported in part by PRESTO Grant Number JPMJPR177A, by Grant-in-Aid for Scientific Research Nos. JP16H05990, JP19H04399 and JP20H05155 from the Japan Society for the Promotion of Science (JSPS), by  MEXT QuantumLeap Flagship Program (MEXT Q-LEAP) Grant Number JPMXS0120184122, by Research Foundation for Opto-Science and Technology, by “Materials research by Information Integration” Initiative (MI$^2$I) project of the Support Program for Starting Up Innovation Hub from JST, the Japan Society for the Promotion of Science through the Funding Program for World-Leading Innovative R\&D on Science and Technology (FIRST Program). V. Ukleev acknowledges support from the SNF Sinergia CRSII5-171003 NanoSkyrmionics. Theoretical part of this study conducted by O. Utesov was funded by RFBR according to the research project 18-32-00083. The research at synchrotron SOLEIL leading to this result has been supported by the project CALIPSOplus under the Grant Agreement 730872 from the EU Framework Programme for Research and Innovation HORIZON 2020. Authors thank A.V. Syromyatnikov, Y. Okamura, J.S. White for the fruitful discussions and Y. Yokoyama and L. Yu for the technical assistance. 

\section*{Appendix A: Conical modulation for the in-plane field $\parallel$ strain}

The quadroupole magnet of the RESOXS setup allowed the in-situ measurement of the magnetic texture evolution for the $B_{\bot}\parallel$ strain geometry (Fig. \ref{fig8}). The upper and the left parts of detector were shadowed by the magnet and the beamstop, respectively. 

The sample was oriented in the beam such as the strain axis was in the horizontal plane and parallel to the magnetic field (Fig. \ref{fig8}a). The RSXS patterns as the function of applied field were measured after ZFC from room temperature to the lowest available temperature $T=32$\,K. The propagation direction of the helical texture after ZFC is similar to the measurements at $T=275$\,K and $T=15$\,K (Fig. \ref{fig2}). Firstly, at the magnetic field of $B_{\bot}=40$\,mT applied parallel to the tensile strain the helical state splits into two domains (Fig. \ref{fig8}b). Next, the two domains merge and the whole texture gradually rotates towards the field direction (Fig. \ref{fig8}c,d). Notably, no higher-harmonic scattering can be observed in the scattering patterns, confirming the smooth helical-to-conical transformation without intermediate CSL state. This observation is in the excellent agreement with our theoretical model. The field-polarized state was not achieved due to the limited amplitude of magnetic field in present setup.

\section*{Appendix B: Multidomain helical state after zero-field cooling and out-of-plane field training}

Zero field cooling from room temperature to $T=15$\,K and subsequent out-of-plane field training at $B_{\parallel}=400$\,mT result to multidomain helical state (Fig. \ref{fig7}). The propagation vectors of the helical modulation are roughly aligned with [111]-equivalent crystallographic axes.


\begin{thebibliography}{59}%
\makeatletter
\providecommand \@ifxundefined [1]{%
 \@ifx{#1\undefined}
}%
\providecommand \@ifnum [1]{%
 \ifnum #1\expandafter \@firstoftwo
 \else \expandafter \@secondoftwo
 \fi
}%
\providecommand \@ifx [1]{%
 \ifx #1\expandafter \@firstoftwo
 \else \expandafter \@secondoftwo
 \fi
}%
\providecommand \natexlab [1]{#1}%
\providecommand \enquote  [1]{``#1''}%
\providecommand \bibnamefont  [1]{#1}%
\providecommand \bibfnamefont [1]{#1}%
\providecommand \citenamefont [1]{#1}%
\providecommand \href@noop [0]{\@secondoftwo}%
\providecommand \href [0]{\begingroup \@sanitize@url \@href}%
\providecommand \@href[1]{\@@startlink{#1}\@@href}%
\providecommand \@@href[1]{\endgroup#1\@@endlink}%
\providecommand \@sanitize@url [0]{\catcode `\\12\catcode `\$12\catcode
  `\&12\catcode `\#12\catcode `\^12\catcode `\_12\catcode `\%12\relax}%
\providecommand \@@startlink[1]{}%
\providecommand \@@endlink[0]{}%
\providecommand \url  [0]{\begingroup\@sanitize@url \@url }%
\providecommand \@url [1]{\endgroup\@href {#1}{\urlprefix }}%
\providecommand \urlprefix  [0]{URL }%
\providecommand \Eprint [0]{\href }%
\providecommand \doibase [0]{http://dx.doi.org/}%
\providecommand \selectlanguage [0]{\@gobble}%
\providecommand \bibinfo  [0]{\@secondoftwo}%
\providecommand \bibfield  [0]{\@secondoftwo}%
\providecommand \translation [1]{[#1]}%
\providecommand \BibitemOpen [0]{}%
\providecommand \bibitemStop [0]{}%
\providecommand \bibitemNoStop [0]{.\EOS\space}%
\providecommand \EOS [0]{\spacefactor3000\relax}%
\providecommand \BibitemShut  [1]{\csname bibitem#1\endcsname}%
\let\auto@bib@innerbib\@empty
\bibitem [{\citenamefont
  {Dzyaloshinsky}(1958)}]{dzyaloshinsky1958thermodynamic}%
  \BibitemOpen
  \bibfield  {author} {\bibinfo {author} {\bibfnamefont {I.}~\bibnamefont
  {Dzyaloshinsky}},\ }\href@noop {} {\bibfield  {journal} {\bibinfo  {journal}
  {Journal of Physics and Chemistry of Solids}\ }\textbf {\bibinfo {volume}
  {4}},\ \bibinfo {pages} {241} (\bibinfo {year} {1958})}\BibitemShut {NoStop}%
\bibitem [{\citenamefont {Moriya}(1960)}]{moriya1960anisotropic}%
  \BibitemOpen
  \bibfield  {author} {\bibinfo {author} {\bibfnamefont {T.}~\bibnamefont
  {Moriya}},\ }\href@noop {} {\bibfield  {journal} {\bibinfo  {journal}
  {Physical Review}\ }\textbf {\bibinfo {volume} {120}},\ \bibinfo {pages} {91}
  (\bibinfo {year} {1960})}\BibitemShut {NoStop}%
\bibitem [{\citenamefont {Bak}\ and\ \citenamefont
  {Jensen}(1980)}]{bak1980theory}%
  \BibitemOpen
  \bibfield  {author} {\bibinfo {author} {\bibfnamefont {P.}~\bibnamefont
  {Bak}}\ and\ \bibinfo {author} {\bibfnamefont {M.~H.}\ \bibnamefont
  {Jensen}},\ }\href@noop {} {\bibfield  {journal} {\bibinfo  {journal}
  {Journal of Physics C: Solid State Physics}\ }\textbf {\bibinfo {volume}
  {13}},\ \bibinfo {pages} {L881} (\bibinfo {year} {1980})}\BibitemShut
  {NoStop}%
\bibitem [{\citenamefont {Dzyaloshinskii}(1965)}]{dzyaloshinskii1965theory}%
  \BibitemOpen
  \bibfield  {author} {\bibinfo {author} {\bibfnamefont {I.}~\bibnamefont
  {Dzyaloshinskii}},\ }\href@noop {} {\bibfield  {journal} {\bibinfo  {journal}
  {Soviet Physics JETP}\ }\textbf {\bibinfo {volume} {20}} (\bibinfo {year}
  {1965})}\BibitemShut {NoStop}%
\bibitem [{\citenamefont {Izyumov}(1984)}]{izyumov1984modulated}%
  \BibitemOpen
  \bibfield  {author} {\bibinfo {author} {\bibfnamefont {Y.~A.}\ \bibnamefont
  {Izyumov}},\ }\href@noop {} {\bibfield  {journal} {\bibinfo  {journal}
  {Physics-Uspekhi}\ }\textbf {\bibinfo {volume} {27}},\ \bibinfo {pages} {845}
  (\bibinfo {year} {1984})}\BibitemShut {NoStop}%
\bibitem [{\citenamefont {Bostrem}\ \emph {et~al.}(2008)\citenamefont
  {Bostrem}, \citenamefont {Kishine},\ and\ \citenamefont
  {Ovchinnikov}}]{bostrem2008theory}%
  \BibitemOpen
  \bibfield  {author} {\bibinfo {author} {\bibfnamefont {I.}~\bibnamefont
  {Bostrem}}, \bibinfo {author} {\bibfnamefont {J.-i.}\ \bibnamefont
  {Kishine}}, \ and\ \bibinfo {author} {\bibfnamefont {A.}~\bibnamefont
  {Ovchinnikov}},\ }\href@noop {} {\bibfield  {journal} {\bibinfo  {journal}
  {Physical Review B}\ }\textbf {\bibinfo {volume} {78}},\ \bibinfo {pages}
  {064425} (\bibinfo {year} {2008})}\BibitemShut {NoStop}%
\bibitem [{\citenamefont {Koumpouras}\ \emph {et~al.}(2016)\citenamefont
  {Koumpouras}, \citenamefont {Bergman}, \citenamefont {Eriksson},\ and\
  \citenamefont {Yudin}}]{koumpouras2016spin}%
  \BibitemOpen
  \bibfield  {author} {\bibinfo {author} {\bibfnamefont {K.}~\bibnamefont
  {Koumpouras}}, \bibinfo {author} {\bibfnamefont {A.}~\bibnamefont {Bergman}},
  \bibinfo {author} {\bibfnamefont {O.}~\bibnamefont {Eriksson}}, \ and\
  \bibinfo {author} {\bibfnamefont {D.}~\bibnamefont {Yudin}},\ }\href@noop {}
  {\bibfield  {journal} {\bibinfo  {journal} {Scientific Reports}\ }\textbf
  {\bibinfo {volume} {6}},\ \bibinfo {pages} {25685} (\bibinfo {year}
  {2016})}\BibitemShut {NoStop}%
\bibitem [{\citenamefont {Togawa}\ \emph {et~al.}(2016)\citenamefont {Togawa},
  \citenamefont {Kousaka}, \citenamefont {Inoue},\ and\ \citenamefont
  {Kishine}}]{togawa2016symmetry}%
  \BibitemOpen
  \bibfield  {author} {\bibinfo {author} {\bibfnamefont {Y.}~\bibnamefont
  {Togawa}}, \bibinfo {author} {\bibfnamefont {Y.}~\bibnamefont {Kousaka}},
  \bibinfo {author} {\bibfnamefont {K.}~\bibnamefont {Inoue}}, \ and\ \bibinfo
  {author} {\bibfnamefont {J.-i.}\ \bibnamefont {Kishine}},\ }\href@noop {}
  {\bibfield  {journal} {\bibinfo  {journal} {Journal of the Physical Society
  of Japan}\ }\textbf {\bibinfo {volume} {85}},\ \bibinfo {pages} {112001}
  (\bibinfo {year} {2016})}\BibitemShut {NoStop}%
\bibitem [{\citenamefont {Goncalves}\ \emph {et~al.}(2017)\citenamefont
  {Goncalves}, \citenamefont {Sogo}, \citenamefont {Shimamoto}, \citenamefont
  {Kousaka}, \citenamefont {Akimitsu}, \citenamefont {Nishihara}, \citenamefont
  {Inoue}, \citenamefont {Yoshizawa}, \citenamefont {Hagiwara}, \citenamefont
  {Mito} \emph {et~al.}}]{goncalves2017collective}%
  \BibitemOpen
  \bibfield  {author} {\bibinfo {author} {\bibfnamefont {F.}~\bibnamefont
  {Goncalves}}, \bibinfo {author} {\bibfnamefont {T.}~\bibnamefont {Sogo}},
  \bibinfo {author} {\bibfnamefont {Y.}~\bibnamefont {Shimamoto}}, \bibinfo
  {author} {\bibfnamefont {Y.}~\bibnamefont {Kousaka}}, \bibinfo {author}
  {\bibfnamefont {J.}~\bibnamefont {Akimitsu}}, \bibinfo {author}
  {\bibfnamefont {S.}~\bibnamefont {Nishihara}}, \bibinfo {author}
  {\bibfnamefont {K.}~\bibnamefont {Inoue}}, \bibinfo {author} {\bibfnamefont
  {D.}~\bibnamefont {Yoshizawa}}, \bibinfo {author} {\bibfnamefont
  {M.}~\bibnamefont {Hagiwara}}, \bibinfo {author} {\bibfnamefont
  {M.}~\bibnamefont {Mito}},  \emph {et~al.},\ }\href@noop {} {\bibfield
  {journal} {\bibinfo  {journal} {Physical Review B}\ }\textbf {\bibinfo
  {volume} {95}},\ \bibinfo {pages} {104415} (\bibinfo {year}
  {2017})}\BibitemShut {NoStop}%
\bibitem [{\citenamefont {Togawa}\ \emph {et~al.}(2012)\citenamefont {Togawa},
  \citenamefont {Koyama}, \citenamefont {Takayanagi}, \citenamefont {Mori},
  \citenamefont {Kousaka}, \citenamefont {Akimitsu}, \citenamefont {Nishihara},
  \citenamefont {Inoue}, \citenamefont {Ovchinnikov},\ and\ \citenamefont
  {Kishine}}]{togawa2012chiral}%
  \BibitemOpen
  \bibfield  {author} {\bibinfo {author} {\bibfnamefont {Y.}~\bibnamefont
  {Togawa}}, \bibinfo {author} {\bibfnamefont {T.}~\bibnamefont {Koyama}},
  \bibinfo {author} {\bibfnamefont {K.}~\bibnamefont {Takayanagi}}, \bibinfo
  {author} {\bibfnamefont {S.}~\bibnamefont {Mori}}, \bibinfo {author}
  {\bibfnamefont {Y.}~\bibnamefont {Kousaka}}, \bibinfo {author} {\bibfnamefont
  {J.}~\bibnamefont {Akimitsu}}, \bibinfo {author} {\bibfnamefont
  {S.}~\bibnamefont {Nishihara}}, \bibinfo {author} {\bibfnamefont
  {K.}~\bibnamefont {Inoue}}, \bibinfo {author} {\bibfnamefont
  {A.}~\bibnamefont {Ovchinnikov}}, \ and\ \bibinfo {author} {\bibfnamefont
  {J.-i.}\ \bibnamefont {Kishine}},\ }\href@noop {} {\bibfield  {journal}
  {\bibinfo  {journal} {Physical Review Letters}\ }\textbf {\bibinfo {volume}
  {108}},\ \bibinfo {pages} {107202} (\bibinfo {year} {2012})}\BibitemShut
  {NoStop}%
\bibitem [{\citenamefont {Togawa}\ \emph {et~al.}(2013)\citenamefont {Togawa},
  \citenamefont {Kousaka}, \citenamefont {Nishihara}, \citenamefont {Inoue},
  \citenamefont {Akimitsu}, \citenamefont {Ovchinnikov},\ and\ \citenamefont
  {Kishine}}]{togawa2013interlayer}%
  \BibitemOpen
  \bibfield  {author} {\bibinfo {author} {\bibfnamefont {Y.}~\bibnamefont
  {Togawa}}, \bibinfo {author} {\bibfnamefont {Y.}~\bibnamefont {Kousaka}},
  \bibinfo {author} {\bibfnamefont {S.}~\bibnamefont {Nishihara}}, \bibinfo
  {author} {\bibfnamefont {K.}~\bibnamefont {Inoue}}, \bibinfo {author}
  {\bibfnamefont {J.}~\bibnamefont {Akimitsu}}, \bibinfo {author}
  {\bibfnamefont {A.}~\bibnamefont {Ovchinnikov}}, \ and\ \bibinfo {author}
  {\bibfnamefont {J.}~\bibnamefont {Kishine}},\ }\href@noop {} {\bibfield
  {journal} {\bibinfo  {journal} {Physical Review Letters}\ }\textbf {\bibinfo
  {volume} {111}},\ \bibinfo {pages} {197204} (\bibinfo {year}
  {2013})}\BibitemShut {NoStop}%
\bibitem [{\citenamefont {Yonemura}\ \emph {et~al.}(2017)\citenamefont
  {Yonemura}, \citenamefont {Shimamoto}, \citenamefont {Kida}, \citenamefont
  {Yoshizawa}, \citenamefont {Kousaka}, \citenamefont {Nishihara},
  \citenamefont {Goncalves}, \citenamefont {Akimitsu}, \citenamefont {Inoue},
  \citenamefont {Hagiwara} \emph {et~al.}}]{yonemura2017magnetic}%
  \BibitemOpen
  \bibfield  {author} {\bibinfo {author} {\bibfnamefont {J.-I.}\ \bibnamefont
  {Yonemura}}, \bibinfo {author} {\bibfnamefont {Y.}~\bibnamefont {Shimamoto}},
  \bibinfo {author} {\bibfnamefont {T.}~\bibnamefont {Kida}}, \bibinfo {author}
  {\bibfnamefont {D.}~\bibnamefont {Yoshizawa}}, \bibinfo {author}
  {\bibfnamefont {Y.}~\bibnamefont {Kousaka}}, \bibinfo {author} {\bibfnamefont
  {S.}~\bibnamefont {Nishihara}}, \bibinfo {author} {\bibfnamefont {F.~J.~T.}\
  \bibnamefont {Goncalves}}, \bibinfo {author} {\bibfnamefont {J.}~\bibnamefont
  {Akimitsu}}, \bibinfo {author} {\bibfnamefont {K.}~\bibnamefont {Inoue}},
  \bibinfo {author} {\bibfnamefont {M.}~\bibnamefont {Hagiwara}},  \emph
  {et~al.},\ }\href@noop {} {\bibfield  {journal} {\bibinfo  {journal}
  {Physical Review B}\ }\textbf {\bibinfo {volume} {96}},\ \bibinfo {pages}
  {184423} (\bibinfo {year} {2017})}\BibitemShut {NoStop}%
\bibitem [{\citenamefont {Tabata}\ \emph {et~al.}(2020)\citenamefont {Tabata},
  \citenamefont {Yamasaki}, \citenamefont {Yokoyama}, \citenamefont {Takagi},
  \citenamefont {Honda}, \citenamefont {Kousaka}, \citenamefont {Akimitsu},\
  and\ \citenamefont {Nakao}}]{tabata2020observation}%
  \BibitemOpen
  \bibfield  {author} {\bibinfo {author} {\bibfnamefont {C.}~\bibnamefont
  {Tabata}}, \bibinfo {author} {\bibfnamefont {Y.}~\bibnamefont {Yamasaki}},
  \bibinfo {author} {\bibfnamefont {Y.}~\bibnamefont {Yokoyama}}, \bibinfo
  {author} {\bibfnamefont {R.}~\bibnamefont {Takagi}}, \bibinfo {author}
  {\bibfnamefont {T.}~\bibnamefont {Honda}}, \bibinfo {author} {\bibfnamefont
  {Y.}~\bibnamefont {Kousaka}}, \bibinfo {author} {\bibfnamefont
  {J.}~\bibnamefont {Akimitsu}}, \ and\ \bibinfo {author} {\bibfnamefont
  {H.}~\bibnamefont {Nakao}},\ }in\ \href@noop {} {\emph {\bibinfo {booktitle}
  {Proceedings of the International Conference on Strongly Correlated Electron
  Systems (SCES2019)}}}\ (\bibinfo {year} {2020})\ p.\ \bibinfo {pages}
  {011194}\BibitemShut {NoStop}%
\bibitem [{\citenamefont {M{\"u}hlbauer}\ \emph {et~al.}(2009)\citenamefont
  {M{\"u}hlbauer}, \citenamefont {Binz}, \citenamefont {Jonietz}, \citenamefont
  {Pfleiderer}, \citenamefont {Rosch}, \citenamefont {Neubauer}, \citenamefont
  {Georgii},\ and\ \citenamefont {B{\"o}ni}}]{muhlbauer2009skyrmion}%
  \BibitemOpen
  \bibfield  {author} {\bibinfo {author} {\bibfnamefont {S.}~\bibnamefont
  {M{\"u}hlbauer}}, \bibinfo {author} {\bibfnamefont {B.}~\bibnamefont {Binz}},
  \bibinfo {author} {\bibfnamefont {F.}~\bibnamefont {Jonietz}}, \bibinfo
  {author} {\bibfnamefont {C.}~\bibnamefont {Pfleiderer}}, \bibinfo {author}
  {\bibfnamefont {A.}~\bibnamefont {Rosch}}, \bibinfo {author} {\bibfnamefont
  {A.}~\bibnamefont {Neubauer}}, \bibinfo {author} {\bibfnamefont
  {R.}~\bibnamefont {Georgii}}, \ and\ \bibinfo {author} {\bibfnamefont
  {P.}~\bibnamefont {B{\"o}ni}},\ }\href@noop {} {\bibfield  {journal}
  {\bibinfo  {journal} {Science}\ }\textbf {\bibinfo {volume} {323}},\ \bibinfo
  {pages} {915} (\bibinfo {year} {2009})}\BibitemShut {NoStop}%
\bibitem [{\citenamefont {Yu}\ \emph {et~al.}(2010)\citenamefont {Yu},
  \citenamefont {Onose}, \citenamefont {Kanazawa}, \citenamefont {Park},
  \citenamefont {Han}, \citenamefont {Matsui}, \citenamefont {Nagaosa},\ and\
  \citenamefont {Tokura}}]{yu2010real}%
  \BibitemOpen
  \bibfield  {author} {\bibinfo {author} {\bibfnamefont {{\relax
  XZ}.}~\bibnamefont {Yu}}, \bibinfo {author} {\bibfnamefont {Y.}~\bibnamefont
  {Onose}}, \bibinfo {author} {\bibfnamefont {N.}~\bibnamefont {Kanazawa}},
  \bibinfo {author} {\bibfnamefont {J.}~\bibnamefont {Park}}, \bibinfo {author}
  {\bibfnamefont {J.}~\bibnamefont {Han}}, \bibinfo {author} {\bibfnamefont
  {Y.}~\bibnamefont {Matsui}}, \bibinfo {author} {\bibfnamefont
  {N.}~\bibnamefont {Nagaosa}}, \ and\ \bibinfo {author} {\bibfnamefont
  {Y.}~\bibnamefont {Tokura}},\ }\href@noop {} {\bibfield  {journal} {\bibinfo
  {journal} {Nature}\ }\textbf {\bibinfo {volume} {465}},\ \bibinfo {pages}
  {901} (\bibinfo {year} {2010})}\BibitemShut {NoStop}%
\bibitem [{\citenamefont {Shibata}\ \emph {et~al.}(2015)\citenamefont
  {Shibata}, \citenamefont {Iwasaki}, \citenamefont {Kanazawa}, \citenamefont
  {Aizawa}, \citenamefont {Tanigaki}, \citenamefont {Shirai}, \citenamefont
  {Nakajima}, \citenamefont {Kubota}, \citenamefont {Kawasaki}, \citenamefont
  {Park} \emph {et~al.}}]{shibata2015large}%
  \BibitemOpen
  \bibfield  {author} {\bibinfo {author} {\bibfnamefont {K.}~\bibnamefont
  {Shibata}}, \bibinfo {author} {\bibfnamefont {J.}~\bibnamefont {Iwasaki}},
  \bibinfo {author} {\bibfnamefont {N.}~\bibnamefont {Kanazawa}}, \bibinfo
  {author} {\bibfnamefont {S.}~\bibnamefont {Aizawa}}, \bibinfo {author}
  {\bibfnamefont {T.}~\bibnamefont {Tanigaki}}, \bibinfo {author}
  {\bibfnamefont {M.}~\bibnamefont {Shirai}}, \bibinfo {author} {\bibfnamefont
  {T.}~\bibnamefont {Nakajima}}, \bibinfo {author} {\bibfnamefont
  {M.}~\bibnamefont {Kubota}}, \bibinfo {author} {\bibfnamefont
  {M.}~\bibnamefont {Kawasaki}}, \bibinfo {author} {\bibfnamefont
  {H.}~\bibnamefont {Park}},  \emph {et~al.},\ }\href@noop {} {\bibfield
  {journal} {\bibinfo  {journal} {Nature Nanotechnology}\ }\textbf {\bibinfo
  {volume} {10}},\ \bibinfo {pages} {589} (\bibinfo {year} {2015})}\BibitemShut
  {NoStop}%
\bibitem [{\citenamefont {Okamura}\ \emph
  {et~al.}(2017{\natexlab{a}})\citenamefont {Okamura}, \citenamefont
  {Yamasaki}, \citenamefont {Morikawa}, \citenamefont {Honda}, \citenamefont
  {Ukleev}, \citenamefont {Nakao}, \citenamefont {Murakami}, \citenamefont
  {Shibata}, \citenamefont {Kagawa}, \citenamefont {Seki} \emph
  {et~al.}}]{okamura2017emergence}%
  \BibitemOpen
  \bibfield  {author} {\bibinfo {author} {\bibfnamefont {Y.}~\bibnamefont
  {Okamura}}, \bibinfo {author} {\bibfnamefont {Y.}~\bibnamefont {Yamasaki}},
  \bibinfo {author} {\bibfnamefont {D.}~\bibnamefont {Morikawa}}, \bibinfo
  {author} {\bibfnamefont {T.}~\bibnamefont {Honda}}, \bibinfo {author}
  {\bibfnamefont {V.}~\bibnamefont {Ukleev}}, \bibinfo {author} {\bibfnamefont
  {H.}~\bibnamefont {Nakao}}, \bibinfo {author} {\bibfnamefont
  {Y.}~\bibnamefont {Murakami}}, \bibinfo {author} {\bibfnamefont
  {K.}~\bibnamefont {Shibata}}, \bibinfo {author} {\bibfnamefont
  {F.}~\bibnamefont {Kagawa}}, \bibinfo {author} {\bibfnamefont
  {S.}~\bibnamefont {Seki}},  \emph {et~al.},\ }\href@noop {} {\bibfield
  {journal} {\bibinfo  {journal} {Physical Review B}\ }\textbf {\bibinfo
  {volume} {96}},\ \bibinfo {pages} {174417} (\bibinfo {year}
  {2017}{\natexlab{a}})}\BibitemShut {NoStop}%
\bibitem [{\citenamefont {Sukhanov}\ \emph {et~al.}(2019)\citenamefont
  {Sukhanov}, \citenamefont {Vir}, \citenamefont {Heinemann}, \citenamefont
  {Nikitin}, \citenamefont {Kriegner}, \citenamefont {Borrmann}, \citenamefont
  {Shekhar}, \citenamefont {Felser},\ and\ \citenamefont
  {Inosov}}]{sukhanov2019giant}%
  \BibitemOpen
  \bibfield  {author} {\bibinfo {author} {\bibfnamefont {A.~S.}\ \bibnamefont
  {Sukhanov}}, \bibinfo {author} {\bibfnamefont {P.}~\bibnamefont {Vir}},
  \bibinfo {author} {\bibfnamefont {A.}~\bibnamefont {Heinemann}}, \bibinfo
  {author} {\bibfnamefont {S.~E.}\ \bibnamefont {Nikitin}}, \bibinfo {author}
  {\bibfnamefont {D.}~\bibnamefont {Kriegner}}, \bibinfo {author}
  {\bibfnamefont {H.}~\bibnamefont {Borrmann}}, \bibinfo {author}
  {\bibfnamefont {C.}~\bibnamefont {Shekhar}}, \bibinfo {author} {\bibfnamefont
  {C.}~\bibnamefont {Felser}}, \ and\ \bibinfo {author} {\bibfnamefont {D.~S.}\
  \bibnamefont {Inosov}},\ }\href {\doibase 10.1103/PhysRevB.100.180403}
  {\bibfield  {journal} {\bibinfo  {journal} {Phys. Rev. B}\ }\textbf {\bibinfo
  {volume} {100}},\ \bibinfo {pages} {180403} (\bibinfo {year}
  {2019})}\BibitemShut {NoStop}%
\bibitem [{\citenamefont {Wilson}\ \emph {et~al.}(2012)\citenamefont {Wilson},
  \citenamefont {Karhu}, \citenamefont {Quigley}, \citenamefont
  {R{\"o}{\ss}ler}, \citenamefont {Butenko}, \citenamefont {Bogdanov},
  \citenamefont {Robertson},\ and\ \citenamefont
  {Monchesky}}]{wilson2012extended}%
  \BibitemOpen
  \bibfield  {author} {\bibinfo {author} {\bibfnamefont {M.}~\bibnamefont
  {Wilson}}, \bibinfo {author} {\bibfnamefont {E.}~\bibnamefont {Karhu}},
  \bibinfo {author} {\bibfnamefont {A.}~\bibnamefont {Quigley}}, \bibinfo
  {author} {\bibfnamefont {U.}~\bibnamefont {R{\"o}{\ss}ler}}, \bibinfo
  {author} {\bibfnamefont {A.}~\bibnamefont {Butenko}}, \bibinfo {author}
  {\bibfnamefont {A.}~\bibnamefont {Bogdanov}}, \bibinfo {author}
  {\bibfnamefont {M.}~\bibnamefont {Robertson}}, \ and\ \bibinfo {author}
  {\bibfnamefont {T.}~\bibnamefont {Monchesky}},\ }\href@noop {} {\bibfield
  {journal} {\bibinfo  {journal} {Physical Review B}\ }\textbf {\bibinfo
  {volume} {86}},\ \bibinfo {pages} {144420} (\bibinfo {year}
  {2012})}\BibitemShut {NoStop}%
\bibitem [{\citenamefont {Huang}\ and\ \citenamefont
  {Chien}(2012)}]{huang2012extended}%
  \BibitemOpen
  \bibfield  {author} {\bibinfo {author} {\bibfnamefont {S.}~\bibnamefont
  {Huang}}\ and\ \bibinfo {author} {\bibfnamefont {C.}~\bibnamefont {Chien}},\
  }\href@noop {} {\bibfield  {journal} {\bibinfo  {journal} {Physical Review
  Letters}\ }\textbf {\bibinfo {volume} {108}},\ \bibinfo {pages} {267201}
  (\bibinfo {year} {2012})}\BibitemShut {NoStop}%
\bibitem [{\citenamefont {Li}\ \emph {et~al.}(2013)\citenamefont {Li},
  \citenamefont {Kanazawa}, \citenamefont {Yu}, \citenamefont {Tsukazaki},
  \citenamefont {Kawasaki}, \citenamefont {Ichikawa}, \citenamefont {Jin},
  \citenamefont {Kagawa},\ and\ \citenamefont {Tokura}}]{li2013robust}%
  \BibitemOpen
  \bibfield  {author} {\bibinfo {author} {\bibfnamefont {Y.}~\bibnamefont
  {Li}}, \bibinfo {author} {\bibfnamefont {N.}~\bibnamefont {Kanazawa}},
  \bibinfo {author} {\bibfnamefont {{\relax XZ}.}~\bibnamefont {Yu}}, \bibinfo
  {author} {\bibfnamefont {A.}~\bibnamefont {Tsukazaki}}, \bibinfo {author}
  {\bibfnamefont {M.}~\bibnamefont {Kawasaki}}, \bibinfo {author}
  {\bibfnamefont {M.}~\bibnamefont {Ichikawa}}, \bibinfo {author}
  {\bibfnamefont {X.}~\bibnamefont {Jin}}, \bibinfo {author} {\bibfnamefont
  {F.}~\bibnamefont {Kagawa}}, \ and\ \bibinfo {author} {\bibfnamefont
  {Y.}~\bibnamefont {Tokura}},\ }\href@noop {} {\bibfield  {journal} {\bibinfo
  {journal} {Physical Review Letters}\ }\textbf {\bibinfo {volume} {110}},\
  \bibinfo {pages} {117202} (\bibinfo {year} {2013})}\BibitemShut {NoStop}%
\bibitem [{\citenamefont {Porter}\ \emph {et~al.}(2015)\citenamefont {Porter},
  \citenamefont {Spencer}, \citenamefont {Temple}, \citenamefont {Kinane},
  \citenamefont {Charlton}, \citenamefont {Langridge},\ and\ \citenamefont
  {Marrows}}]{porter2015manipulation}%
  \BibitemOpen
  \bibfield  {author} {\bibinfo {author} {\bibfnamefont {N.~A.}\ \bibnamefont
  {Porter}}, \bibinfo {author} {\bibfnamefont {C.~S.}\ \bibnamefont {Spencer}},
  \bibinfo {author} {\bibfnamefont {R.~C.}\ \bibnamefont {Temple}}, \bibinfo
  {author} {\bibfnamefont {C.~J.}\ \bibnamefont {Kinane}}, \bibinfo {author}
  {\bibfnamefont {T.~R.}\ \bibnamefont {Charlton}}, \bibinfo {author}
  {\bibfnamefont {S.}~\bibnamefont {Langridge}}, \ and\ \bibinfo {author}
  {\bibfnamefont {C.~H.}\ \bibnamefont {Marrows}},\ }\href@noop {} {\bibfield
  {journal} {\bibinfo  {journal} {Physical Review B}\ }\textbf {\bibinfo
  {volume} {92}},\ \bibinfo {pages} {144402} (\bibinfo {year}
  {2015})}\BibitemShut {NoStop}%
\bibitem [{\citenamefont {Kanazawa}\ \emph {et~al.}(2016)\citenamefont
  {Kanazawa}, \citenamefont {White}, \citenamefont {R{\o}nnow}, \citenamefont
  {Dewhurst}, \citenamefont {Fujishiro}, \citenamefont {Tsukazaki},
  \citenamefont {Kozuka}, \citenamefont {Kawasaki}, \citenamefont {Ichikawa},
  \citenamefont {Kagawa} \emph {et~al.}}]{kanazawa2016direct}%
  \BibitemOpen
  \bibfield  {author} {\bibinfo {author} {\bibfnamefont {N.}~\bibnamefont
  {Kanazawa}}, \bibinfo {author} {\bibfnamefont {J.}~\bibnamefont {White}},
  \bibinfo {author} {\bibfnamefont {H.~M.}\ \bibnamefont {R{\o}nnow}}, \bibinfo
  {author} {\bibfnamefont {C.}~\bibnamefont {Dewhurst}}, \bibinfo {author}
  {\bibfnamefont {Y.}~\bibnamefont {Fujishiro}}, \bibinfo {author}
  {\bibfnamefont {A.}~\bibnamefont {Tsukazaki}}, \bibinfo {author}
  {\bibfnamefont {Y.}~\bibnamefont {Kozuka}}, \bibinfo {author} {\bibfnamefont
  {M.}~\bibnamefont {Kawasaki}}, \bibinfo {author} {\bibfnamefont
  {M.}~\bibnamefont {Ichikawa}}, \bibinfo {author} {\bibfnamefont
  {F.}~\bibnamefont {Kagawa}},  \emph {et~al.},\ }\href@noop {} {\bibfield
  {journal} {\bibinfo  {journal} {Physical Review B}\ }\textbf {\bibinfo
  {volume} {94}},\ \bibinfo {pages} {184432} (\bibinfo {year}
  {2016})}\BibitemShut {NoStop}%
\bibitem [{\citenamefont {Zhang}\ \emph {et~al.}(2017)\citenamefont {Zhang},
  \citenamefont {Stasinopoulos}, \citenamefont {Lancaster}, \citenamefont
  {Xiao}, \citenamefont {Bauer}, \citenamefont {Rucker}, \citenamefont {Baker},
  \citenamefont {Figueroa}, \citenamefont {Salman}, \citenamefont {Pratt} \emph
  {et~al.}}]{zhang2017room}%
  \BibitemOpen
  \bibfield  {author} {\bibinfo {author} {\bibfnamefont {S.}~\bibnamefont
  {Zhang}}, \bibinfo {author} {\bibfnamefont {I.}~\bibnamefont
  {Stasinopoulos}}, \bibinfo {author} {\bibfnamefont {T.}~\bibnamefont
  {Lancaster}}, \bibinfo {author} {\bibfnamefont {F.}~\bibnamefont {Xiao}},
  \bibinfo {author} {\bibfnamefont {A.}~\bibnamefont {Bauer}}, \bibinfo
  {author} {\bibfnamefont {F.}~\bibnamefont {Rucker}}, \bibinfo {author}
  {\bibfnamefont {A.}~\bibnamefont {Baker}}, \bibinfo {author} {\bibfnamefont
  {A.}~\bibnamefont {Figueroa}}, \bibinfo {author} {\bibfnamefont
  {Z.}~\bibnamefont {Salman}}, \bibinfo {author} {\bibfnamefont
  {F.}~\bibnamefont {Pratt}},  \emph {et~al.},\ }\href@noop {} {\bibfield
  {journal} {\bibinfo  {journal} {Scientific Reports}\ }\textbf {\bibinfo
  {volume} {7}},\ \bibinfo {pages} {123} (\bibinfo {year} {2017})}\BibitemShut
  {NoStop}%
\bibitem [{\citenamefont {Oike}\ \emph {et~al.}(2016)\citenamefont {Oike},
  \citenamefont {Kikkawa}, \citenamefont {Kanazawa}, \citenamefont {Taguchi},
  \citenamefont {Kawasaki}, \citenamefont {Tokura},\ and\ \citenamefont
  {Kagawa}}]{oike2016interplay}%
  \BibitemOpen
  \bibfield  {author} {\bibinfo {author} {\bibfnamefont {H.}~\bibnamefont
  {Oike}}, \bibinfo {author} {\bibfnamefont {A.}~\bibnamefont {Kikkawa}},
  \bibinfo {author} {\bibfnamefont {N.}~\bibnamefont {Kanazawa}}, \bibinfo
  {author} {\bibfnamefont {Y.}~\bibnamefont {Taguchi}}, \bibinfo {author}
  {\bibfnamefont {M.}~\bibnamefont {Kawasaki}}, \bibinfo {author}
  {\bibfnamefont {Y.}~\bibnamefont {Tokura}}, \ and\ \bibinfo {author}
  {\bibfnamefont {F.}~\bibnamefont {Kagawa}},\ }\href@noop {} {\bibfield
  {journal} {\bibinfo  {journal} {Nature Physics}\ }\textbf {\bibinfo {volume}
  {12}},\ \bibinfo {pages} {62} (\bibinfo {year} {2016})}\BibitemShut {NoStop}%
\bibitem [{\citenamefont {Karube}\ \emph {et~al.}(2017)\citenamefont {Karube},
  \citenamefont {White}, \citenamefont {Morikawa}, \citenamefont {Bartkowiak},
  \citenamefont {Kikkawa}, \citenamefont {Tokunaga}, \citenamefont {Arima},
  \citenamefont {R{\o}nnow}, \citenamefont {Tokura},\ and\ \citenamefont
  {Taguchi}}]{karube2017skyrmion}%
  \BibitemOpen
  \bibfield  {author} {\bibinfo {author} {\bibfnamefont {K.}~\bibnamefont
  {Karube}}, \bibinfo {author} {\bibfnamefont {J.}~\bibnamefont {White}},
  \bibinfo {author} {\bibfnamefont {D.}~\bibnamefont {Morikawa}}, \bibinfo
  {author} {\bibfnamefont {M.}~\bibnamefont {Bartkowiak}}, \bibinfo {author}
  {\bibfnamefont {A.}~\bibnamefont {Kikkawa}}, \bibinfo {author} {\bibfnamefont
  {Y.}~\bibnamefont {Tokunaga}}, \bibinfo {author} {\bibfnamefont {T.-H.}\
  \bibnamefont {Arima}}, \bibinfo {author} {\bibfnamefont {H.}~\bibnamefont
  {R{\o}nnow}}, \bibinfo {author} {\bibfnamefont {Y.}~\bibnamefont {Tokura}}, \
  and\ \bibinfo {author} {\bibfnamefont {Y.}~\bibnamefont {Taguchi}},\
  }\href@noop {} {\bibfield  {journal} {\bibinfo  {journal} {Physical Review
  Materials}\ }\textbf {\bibinfo {volume} {1}},\ \bibinfo {pages} {074405}
  (\bibinfo {year} {2017})}\BibitemShut {NoStop}%
\bibitem [{\citenamefont {Yu}\ \emph {et~al.}(2018)\citenamefont {Yu},
  \citenamefont {Morikawa}, \citenamefont {Yokouchi}, \citenamefont {Shibata},
  \citenamefont {Kanazawa}, \citenamefont {Kagawa}, \citenamefont {Arima},\
  and\ \citenamefont {Tokura}}]{yu2018aggregation}%
  \BibitemOpen
  \bibfield  {author} {\bibinfo {author} {\bibfnamefont {{\relax
  XZ}.}~\bibnamefont {Yu}}, \bibinfo {author} {\bibfnamefont {D.}~\bibnamefont
  {Morikawa}}, \bibinfo {author} {\bibfnamefont {T.}~\bibnamefont {Yokouchi}},
  \bibinfo {author} {\bibfnamefont {K.}~\bibnamefont {Shibata}}, \bibinfo
  {author} {\bibfnamefont {N.}~\bibnamefont {Kanazawa}}, \bibinfo {author}
  {\bibfnamefont {F.}~\bibnamefont {Kagawa}}, \bibinfo {author} {\bibfnamefont
  {T.-H.}\ \bibnamefont {Arima}}, \ and\ \bibinfo {author} {\bibfnamefont
  {Y.}~\bibnamefont {Tokura}},\ }\href@noop {} {\bibfield  {journal} {\bibinfo
  {journal} {Nature Physics}\ ,\ \bibinfo {pages} {1}} (\bibinfo {year}
  {2018})}\BibitemShut {NoStop}%
\bibitem [{\citenamefont {Nakajima}\ \emph {et~al.}(2018)\citenamefont
  {Nakajima}, \citenamefont {Ukleev}, \citenamefont {Ohishi}, \citenamefont
  {Oike}, \citenamefont {Kagawa}, \citenamefont {Seki}, \citenamefont
  {Kakurai}, \citenamefont {Tokura},\ and\ \citenamefont
  {Arima}}]{nakajima2018uniaxial}%
  \BibitemOpen
  \bibfield  {author} {\bibinfo {author} {\bibfnamefont {T.}~\bibnamefont
  {Nakajima}}, \bibinfo {author} {\bibfnamefont {V.}~\bibnamefont {Ukleev}},
  \bibinfo {author} {\bibfnamefont {K.}~\bibnamefont {Ohishi}}, \bibinfo
  {author} {\bibfnamefont {H.}~\bibnamefont {Oike}}, \bibinfo {author}
  {\bibfnamefont {F.}~\bibnamefont {Kagawa}}, \bibinfo {author} {\bibfnamefont
  {S.-I.}\ \bibnamefont {Seki}}, \bibinfo {author} {\bibfnamefont
  {K.}~\bibnamefont {Kakurai}}, \bibinfo {author} {\bibfnamefont
  {Y.}~\bibnamefont {Tokura}}, \ and\ \bibinfo {author} {\bibfnamefont {T.-H.}\
  \bibnamefont {Arima}},\ }\href@noop {} {\bibfield  {journal} {\bibinfo
  {journal} {Journal of the Physical Society of Japan}\ }\textbf {\bibinfo
  {volume} {87}},\ \bibinfo {pages} {094709} (\bibinfo {year}
  {2018})}\BibitemShut {NoStop}%
\bibitem [{\citenamefont {Qian}\ \emph {et~al.}(2018)\citenamefont {Qian},
  \citenamefont {Bannenberg}, \citenamefont {Wilhelm}, \citenamefont
  {Chaboussant}, \citenamefont {Debeer-Schmitt}, \citenamefont {Schmidt},
  \citenamefont {Aqeel}, \citenamefont {Palstra}, \citenamefont {Br{\"u}ck},
  \citenamefont {Lefering} \emph {et~al.}}]{qian2018new}%
  \BibitemOpen
  \bibfield  {author} {\bibinfo {author} {\bibfnamefont {F.}~\bibnamefont
  {Qian}}, \bibinfo {author} {\bibfnamefont {L.~J.}\ \bibnamefont
  {Bannenberg}}, \bibinfo {author} {\bibfnamefont {H.}~\bibnamefont {Wilhelm}},
  \bibinfo {author} {\bibfnamefont {G.}~\bibnamefont {Chaboussant}}, \bibinfo
  {author} {\bibfnamefont {L.~M.}\ \bibnamefont {Debeer-Schmitt}}, \bibinfo
  {author} {\bibfnamefont {M.~P.}\ \bibnamefont {Schmidt}}, \bibinfo {author}
  {\bibfnamefont {A.}~\bibnamefont {Aqeel}}, \bibinfo {author} {\bibfnamefont
  {T.~T.}\ \bibnamefont {Palstra}}, \bibinfo {author} {\bibfnamefont
  {E.}~\bibnamefont {Br{\"u}ck}}, \bibinfo {author} {\bibfnamefont {A.~J.}\
  \bibnamefont {Lefering}},  \emph {et~al.},\ }\href@noop {} {\bibfield
  {journal} {\bibinfo  {journal} {Science Advances}\ }\textbf {\bibinfo
  {volume} {4}},\ \bibinfo {pages} {eaat7323} (\bibinfo {year}
  {2018})}\BibitemShut {NoStop}%
\bibitem [{\citenamefont {White}\ \emph {et~al.}(2018)\citenamefont {White},
  \citenamefont {{\v{Z}}ivkovi{\'c}}, \citenamefont {Kruchkov}, \citenamefont
  {Bartkowiak}, \citenamefont {Magrez},\ and\ \citenamefont
  {R{\o}nnow}}]{white2018electric}%
  \BibitemOpen
  \bibfield  {author} {\bibinfo {author} {\bibfnamefont {J.}~\bibnamefont
  {White}}, \bibinfo {author} {\bibfnamefont {I.}~\bibnamefont
  {{\v{Z}}ivkovi{\'c}}}, \bibinfo {author} {\bibfnamefont {A.}~\bibnamefont
  {Kruchkov}}, \bibinfo {author} {\bibfnamefont {M.}~\bibnamefont
  {Bartkowiak}}, \bibinfo {author} {\bibfnamefont {A.}~\bibnamefont {Magrez}},
  \ and\ \bibinfo {author} {\bibfnamefont {H.}~\bibnamefont {R{\o}nnow}},\
  }\href@noop {} {\bibfield  {journal} {\bibinfo  {journal} {Physical Review
  Applied}\ }\textbf {\bibinfo {volume} {10}},\ \bibinfo {pages} {014021}
  (\bibinfo {year} {2018})}\BibitemShut {NoStop}%
\bibitem [{\citenamefont {Chacon}\ \emph {et~al.}(2018)\citenamefont {Chacon},
  \citenamefont {Heinen}, \citenamefont {Halder}, \citenamefont {Bauer},
  \citenamefont {Simeth}, \citenamefont {M{\"u}hlbauer}, \citenamefont
  {Berger}, \citenamefont {Garst}, \citenamefont {Rosch},\ and\ \citenamefont
  {Pfleiderer}}]{chacon2018observation}%
  \BibitemOpen
  \bibfield  {author} {\bibinfo {author} {\bibfnamefont {A.}~\bibnamefont
  {Chacon}}, \bibinfo {author} {\bibfnamefont {L.}~\bibnamefont {Heinen}},
  \bibinfo {author} {\bibfnamefont {M.}~\bibnamefont {Halder}}, \bibinfo
  {author} {\bibfnamefont {A.}~\bibnamefont {Bauer}}, \bibinfo {author}
  {\bibfnamefont {W.}~\bibnamefont {Simeth}}, \bibinfo {author} {\bibfnamefont
  {S.}~\bibnamefont {M{\"u}hlbauer}}, \bibinfo {author} {\bibfnamefont
  {H.}~\bibnamefont {Berger}}, \bibinfo {author} {\bibfnamefont
  {M.}~\bibnamefont {Garst}}, \bibinfo {author} {\bibfnamefont
  {A.}~\bibnamefont {Rosch}}, \ and\ \bibinfo {author} {\bibfnamefont
  {C.}~\bibnamefont {Pfleiderer}},\ }\href@noop {} {\bibfield  {journal}
  {\bibinfo  {journal} {Nature Physics}\ }\textbf {\bibinfo {volume} {14}},\
  \bibinfo {pages} {936} (\bibinfo {year} {2018})}\BibitemShut {NoStop}%
\bibitem [{\citenamefont {Wilson}\ \emph {et~al.}(2020)\citenamefont {Wilson},
  \citenamefont {Birch}, \citenamefont {{\v{S}}tefan{\v{c}}i{\v{c}}},
  \citenamefont {Twitchett-Harrison}, \citenamefont {Balakrishnan},
  \citenamefont {Hicken}, \citenamefont {Fan}, \citenamefont {Steadman},\ and\
  \citenamefont {Hatton}}]{wilson2020stability}%
  \BibitemOpen
  \bibfield  {author} {\bibinfo {author} {\bibfnamefont {M.}~\bibnamefont
  {Wilson}}, \bibinfo {author} {\bibfnamefont {M.}~\bibnamefont {Birch}},
  \bibinfo {author} {\bibfnamefont {A.}~\bibnamefont
  {{\v{S}}tefan{\v{c}}i{\v{c}}}}, \bibinfo {author} {\bibfnamefont
  {A.}~\bibnamefont {Twitchett-Harrison}}, \bibinfo {author} {\bibfnamefont
  {G.}~\bibnamefont {Balakrishnan}}, \bibinfo {author} {\bibfnamefont
  {T.}~\bibnamefont {Hicken}}, \bibinfo {author} {\bibfnamefont
  {R.}~\bibnamefont {Fan}}, \bibinfo {author} {\bibfnamefont {P.}~\bibnamefont
  {Steadman}}, \ and\ \bibinfo {author} {\bibfnamefont {P.}~\bibnamefont
  {Hatton}},\ }\href@noop {} {\bibfield  {journal} {\bibinfo  {journal}
  {Physical Review Research}\ }\textbf {\bibinfo {volume} {2}},\ \bibinfo
  {pages} {013096} (\bibinfo {year} {2020})}\BibitemShut {NoStop}%
\bibitem [{\citenamefont {Yamasaki}\ \emph {et~al.}(2015)\citenamefont
  {Yamasaki}, \citenamefont {Morikawa}, \citenamefont {Honda}, \citenamefont
  {Nakao}, \citenamefont {Murakami}, \citenamefont {Kanazawa}, \citenamefont
  {Kawasaki}, \citenamefont {Arima},\ and\ \citenamefont
  {Tokura}}]{yamasaki2015dynamical}%
  \BibitemOpen
  \bibfield  {author} {\bibinfo {author} {\bibfnamefont {Y.}~\bibnamefont
  {Yamasaki}}, \bibinfo {author} {\bibfnamefont {D.}~\bibnamefont {Morikawa}},
  \bibinfo {author} {\bibfnamefont {T.}~\bibnamefont {Honda}}, \bibinfo
  {author} {\bibfnamefont {H.}~\bibnamefont {Nakao}}, \bibinfo {author}
  {\bibfnamefont {Y.}~\bibnamefont {Murakami}}, \bibinfo {author}
  {\bibfnamefont {N.}~\bibnamefont {Kanazawa}}, \bibinfo {author}
  {\bibfnamefont {M.}~\bibnamefont {Kawasaki}}, \bibinfo {author}
  {\bibfnamefont {T.-H.}\ \bibnamefont {Arima}}, \ and\ \bibinfo {author}
  {\bibfnamefont {Y.}~\bibnamefont {Tokura}},\ }\href@noop {} {\bibfield
  {journal} {\bibinfo  {journal} {Physical Review B}\ }\textbf {\bibinfo
  {volume} {92}},\ \bibinfo {pages} {220421} (\bibinfo {year}
  {2015})}\BibitemShut {NoStop}%
\bibitem [{\citenamefont {Jaouen}\ \emph {et~al.}(2004)\citenamefont {Jaouen},
  \citenamefont {Tonnerre}, \citenamefont {Kapoujian}, \citenamefont {Taunier},
  \citenamefont {Roux}, \citenamefont {Raoux},\ and\ \citenamefont
  {Sirotti}}]{jaouen2004apparatus}%
  \BibitemOpen
  \bibfield  {author} {\bibinfo {author} {\bibfnamefont {N.}~\bibnamefont
  {Jaouen}}, \bibinfo {author} {\bibfnamefont {J.-M.}\ \bibnamefont
  {Tonnerre}}, \bibinfo {author} {\bibfnamefont {G.}~\bibnamefont {Kapoujian}},
  \bibinfo {author} {\bibfnamefont {P.}~\bibnamefont {Taunier}}, \bibinfo
  {author} {\bibfnamefont {J.-P.}\ \bibnamefont {Roux}}, \bibinfo {author}
  {\bibfnamefont {D.}~\bibnamefont {Raoux}}, \ and\ \bibinfo {author}
  {\bibfnamefont {F.}~\bibnamefont {Sirotti}},\ }\href@noop {} {\bibfield
  {journal} {\bibinfo  {journal} {Journal of Synchrotron Radiation}\ }\textbf
  {\bibinfo {volume} {11}},\ \bibinfo {pages} {353} (\bibinfo {year}
  {2004})}\BibitemShut {NoStop}%
\bibitem [{\citenamefont {Kanazawa}(2015)}]{kanazawa2015magnetic}%
  \BibitemOpen
  \bibfield  {author} {\bibinfo {author} {\bibfnamefont {N.}~\bibnamefont
  {Kanazawa}},\ }\enquote {\bibinfo {title} {Magnetic and transport properties
  in b20-type germanides},}\ in\ \href {\doibase 10.1007/978-4-431-55660-2_3}
  {\emph {\bibinfo {booktitle} {Charge and Heat Transport Phenomena in
  Electronic and Spin Structures in B20-type Compounds}}}\ (\bibinfo
  {publisher} {Springer Japan},\ \bibinfo {address} {Tokyo},\ \bibinfo {year}
  {2015})\ pp.\ \bibinfo {pages} {29--44}\BibitemShut {NoStop}%
\bibitem [{\citenamefont {Ukleev}\ \emph {et~al.}(2018)\citenamefont {Ukleev},
  \citenamefont {Yamasaki}, \citenamefont {Morikawa}, \citenamefont {Kanazawa},
  \citenamefont {Okamura}, \citenamefont {Nakao}, \citenamefont {Tokura} \emph
  {et~al.}}]{ukleev2018coherent}%
  \BibitemOpen
  \bibfield  {author} {\bibinfo {author} {\bibfnamefont {V.}~\bibnamefont
  {Ukleev}}, \bibinfo {author} {\bibfnamefont {Y.}~\bibnamefont {Yamasaki}},
  \bibinfo {author} {\bibfnamefont {D.}~\bibnamefont {Morikawa}}, \bibinfo
  {author} {\bibfnamefont {N.}~\bibnamefont {Kanazawa}}, \bibinfo {author}
  {\bibfnamefont {Y.}~\bibnamefont {Okamura}}, \bibinfo {author} {\bibfnamefont
  {H.}~\bibnamefont {Nakao}}, \bibinfo {author} {\bibfnamefont
  {Y.}~\bibnamefont {Tokura}},  \emph {et~al.},\ }\href@noop {} {\bibfield
  {journal} {\bibinfo  {journal} {Quantum Beam Science}\ }\textbf {\bibinfo
  {volume} {2}},\ \bibinfo {pages} {3} (\bibinfo {year} {2018})}\BibitemShut
  {NoStop}%
\bibitem [{\citenamefont {Burn}\ \emph {et~al.}(2019)\citenamefont {Burn},
  \citenamefont {Zhang}, \citenamefont {Wang}, \citenamefont {Du},
  \citenamefont {Van Der~Laan},\ and\ \citenamefont
  {Hesjedal}}]{burn2019helical}%
  \BibitemOpen
  \bibfield  {author} {\bibinfo {author} {\bibfnamefont {D.}~\bibnamefont
  {Burn}}, \bibinfo {author} {\bibfnamefont {S.}~\bibnamefont {Zhang}},
  \bibinfo {author} {\bibfnamefont {S.}~\bibnamefont {Wang}}, \bibinfo {author}
  {\bibfnamefont {H.}~\bibnamefont {Du}}, \bibinfo {author} {\bibfnamefont
  {G.}~\bibnamefont {Van Der~Laan}}, \ and\ \bibinfo {author} {\bibfnamefont
  {T.}~\bibnamefont {Hesjedal}},\ }\href@noop {} {\bibfield  {journal}
  {\bibinfo  {journal} {Physical Review B}\ }\textbf {\bibinfo {volume}
  {100}},\ \bibinfo {pages} {184403} (\bibinfo {year} {2019})}\BibitemShut
  {NoStop}%
\bibitem [{\citenamefont {Burn}\ \emph {et~al.}(2020)\citenamefont {Burn},
  \citenamefont {Wang}, \citenamefont {Wang}, \citenamefont {van~der Laan},
  \citenamefont {Zhang}, \citenamefont {Du},\ and\ \citenamefont
  {Hesjedal}}]{burn2020field}%
  \BibitemOpen
  \bibfield  {author} {\bibinfo {author} {\bibfnamefont {D.~M.}\ \bibnamefont
  {Burn}}, \bibinfo {author} {\bibfnamefont {S.}~\bibnamefont {Wang}}, \bibinfo
  {author} {\bibfnamefont {W.}~\bibnamefont {Wang}}, \bibinfo {author}
  {\bibfnamefont {G.}~\bibnamefont {van~der Laan}}, \bibinfo {author}
  {\bibfnamefont {S.}~\bibnamefont {Zhang}}, \bibinfo {author} {\bibfnamefont
  {H.}~\bibnamefont {Du}}, \ and\ \bibinfo {author} {\bibfnamefont
  {T.}~\bibnamefont {Hesjedal}},\ }\href@noop {} {\bibfield  {journal}
  {\bibinfo  {journal} {Physical Review B}\ }\textbf {\bibinfo {volume}
  {101}},\ \bibinfo {pages} {014446} (\bibinfo {year} {2020})}\BibitemShut
  {NoStop}%
\bibitem [{\citenamefont {Okamura}\ \emph
  {et~al.}(2017{\natexlab{b}})\citenamefont {Okamura}, \citenamefont
  {Yamasaki}, \citenamefont {Morikawa}, \citenamefont {Honda}, \citenamefont
  {Ukleev}, \citenamefont {Nakao}, \citenamefont {Murakami}, \citenamefont
  {Shibata}, \citenamefont {Kagawa}, \citenamefont {Seki} \emph
  {et~al.}}]{okamura2017directional}%
  \BibitemOpen
  \bibfield  {author} {\bibinfo {author} {\bibfnamefont {Y.}~\bibnamefont
  {Okamura}}, \bibinfo {author} {\bibfnamefont {Y.}~\bibnamefont {Yamasaki}},
  \bibinfo {author} {\bibfnamefont {D.}~\bibnamefont {Morikawa}}, \bibinfo
  {author} {\bibfnamefont {T.}~\bibnamefont {Honda}}, \bibinfo {author}
  {\bibfnamefont {V.}~\bibnamefont {Ukleev}}, \bibinfo {author} {\bibfnamefont
  {H.}~\bibnamefont {Nakao}}, \bibinfo {author} {\bibfnamefont
  {Y.}~\bibnamefont {Murakami}}, \bibinfo {author} {\bibfnamefont
  {K.}~\bibnamefont {Shibata}}, \bibinfo {author} {\bibfnamefont
  {F.}~\bibnamefont {Kagawa}}, \bibinfo {author} {\bibfnamefont
  {S.}~\bibnamefont {Seki}},  \emph {et~al.},\ }\href@noop {} {\bibfield
  {journal} {\bibinfo  {journal} {Physical Review B}\ }\textbf {\bibinfo
  {volume} {95}},\ \bibinfo {pages} {184411} (\bibinfo {year}
  {2017}{\natexlab{b}})}\BibitemShut {NoStop}%
\bibitem [{\citenamefont {Grigoriev}\ \emph {et~al.}(2006)\citenamefont
  {Grigoriev}, \citenamefont {Maleyev}, \citenamefont {Okorokov}, \citenamefont
  {Chetverikov}, \citenamefont {B{\"o}ni}, \citenamefont {Georgii},
  \citenamefont {Lamago}, \citenamefont {Eckerlebe},\ and\ \citenamefont
  {Pranzas}}]{grigoriev2006magnetic}%
  \BibitemOpen
  \bibfield  {author} {\bibinfo {author} {\bibfnamefont {S.}~\bibnamefont
  {Grigoriev}}, \bibinfo {author} {\bibfnamefont {S.}~\bibnamefont {Maleyev}},
  \bibinfo {author} {\bibfnamefont {A.}~\bibnamefont {Okorokov}}, \bibinfo
  {author} {\bibfnamefont {Y.~O.}\ \bibnamefont {Chetverikov}}, \bibinfo
  {author} {\bibfnamefont {P.}~\bibnamefont {B{\"o}ni}}, \bibinfo {author}
  {\bibfnamefont {R.}~\bibnamefont {Georgii}}, \bibinfo {author} {\bibfnamefont
  {D.}~\bibnamefont {Lamago}}, \bibinfo {author} {\bibfnamefont
  {H.}~\bibnamefont {Eckerlebe}}, \ and\ \bibinfo {author} {\bibfnamefont
  {K.}~\bibnamefont {Pranzas}},\ }\href@noop {} {\bibfield  {journal} {\bibinfo
   {journal} {Physical Review B}\ }\textbf {\bibinfo {volume} {74}},\ \bibinfo
  {pages} {214414} (\bibinfo {year} {2006})}\BibitemShut {NoStop}%
\bibitem [{\citenamefont {D{\"u}rr}\ \emph {et~al.}(1999)\citenamefont
  {D{\"u}rr}, \citenamefont {Dudzik}, \citenamefont {Dhesi}, \citenamefont
  {Goedkoop}, \citenamefont {Van~der Laan}, \citenamefont {Belakhovsky},
  \citenamefont {Mocuta}, \citenamefont {Marty},\ and\ \citenamefont
  {Samson}}]{durr1999chiral}%
  \BibitemOpen
  \bibfield  {author} {\bibinfo {author} {\bibfnamefont {H.}~\bibnamefont
  {D{\"u}rr}}, \bibinfo {author} {\bibfnamefont {E.}~\bibnamefont {Dudzik}},
  \bibinfo {author} {\bibfnamefont {S.}~\bibnamefont {Dhesi}}, \bibinfo
  {author} {\bibfnamefont {J.}~\bibnamefont {Goedkoop}}, \bibinfo {author}
  {\bibfnamefont {G.}~\bibnamefont {Van~der Laan}}, \bibinfo {author}
  {\bibfnamefont {M.}~\bibnamefont {Belakhovsky}}, \bibinfo {author}
  {\bibfnamefont {C.}~\bibnamefont {Mocuta}}, \bibinfo {author} {\bibfnamefont
  {A.}~\bibnamefont {Marty}}, \ and\ \bibinfo {author} {\bibfnamefont
  {Y.}~\bibnamefont {Samson}},\ }\href@noop {} {\bibfield  {journal} {\bibinfo
  {journal} {Science}\ }\textbf {\bibinfo {volume} {284}},\ \bibinfo {pages}
  {2166} (\bibinfo {year} {1999})}\BibitemShut {NoStop}%
\bibitem [{\citenamefont {Hellwig}\ \emph {et~al.}(2003)\citenamefont
  {Hellwig}, \citenamefont {Denbeaux}, \citenamefont {Kortright},\ and\
  \citenamefont {Fullerton}}]{hellwig2003x}%
  \BibitemOpen
  \bibfield  {author} {\bibinfo {author} {\bibfnamefont {O.}~\bibnamefont
  {Hellwig}}, \bibinfo {author} {\bibfnamefont {G.}~\bibnamefont {Denbeaux}},
  \bibinfo {author} {\bibfnamefont {J.}~\bibnamefont {Kortright}}, \ and\
  \bibinfo {author} {\bibfnamefont {E.~E.}\ \bibnamefont {Fullerton}},\
  }\href@noop {} {\bibfield  {journal} {\bibinfo  {journal} {Physica B:
  Condensed Matter}\ }\textbf {\bibinfo {volume} {336}},\ \bibinfo {pages}
  {136} (\bibinfo {year} {2003})}\BibitemShut {NoStop}%
\bibitem [{\citenamefont {Desautels}\ \emph {et~al.}(2019)\citenamefont
  {Desautels}, \citenamefont {DeBeer-Schmitt}, \citenamefont {Montoya},
  \citenamefont {Borchers}, \citenamefont {Je}, \citenamefont {Tang},
  \citenamefont {Im}, \citenamefont {Fitzsimmons}, \citenamefont {Fullerton},\
  and\ \citenamefont {Gilbert}}]{desautels2019realization}%
  \BibitemOpen
  \bibfield  {author} {\bibinfo {author} {\bibfnamefont {R.~D.}\ \bibnamefont
  {Desautels}}, \bibinfo {author} {\bibfnamefont {L.}~\bibnamefont
  {DeBeer-Schmitt}}, \bibinfo {author} {\bibfnamefont {S.~A.}\ \bibnamefont
  {Montoya}}, \bibinfo {author} {\bibfnamefont {J.~A.}\ \bibnamefont
  {Borchers}}, \bibinfo {author} {\bibfnamefont {S.-G.}\ \bibnamefont {Je}},
  \bibinfo {author} {\bibfnamefont {N.}~\bibnamefont {Tang}}, \bibinfo {author}
  {\bibfnamefont {M.-Y.}\ \bibnamefont {Im}}, \bibinfo {author} {\bibfnamefont
  {M.~R.}\ \bibnamefont {Fitzsimmons}}, \bibinfo {author} {\bibfnamefont
  {E.~E.}\ \bibnamefont {Fullerton}}, \ and\ \bibinfo {author} {\bibfnamefont
  {D.~A.}\ \bibnamefont {Gilbert}},\ }\href@noop {} {\bibfield  {journal}
  {\bibinfo  {journal} {Physical Review Materials}\ }\textbf {\bibinfo {volume}
  {3}},\ \bibinfo {pages} {104406} (\bibinfo {year} {2019})}\BibitemShut
  {NoStop}%
\bibitem [{\citenamefont {Bogdanov}\ and\ \citenamefont
  {Hubert}(1994)}]{bogdanov1994thermodynamically}%
  \BibitemOpen
  \bibfield  {author} {\bibinfo {author} {\bibfnamefont {A.}~\bibnamefont
  {Bogdanov}}\ and\ \bibinfo {author} {\bibfnamefont {A.}~\bibnamefont
  {Hubert}},\ }\href@noop {} {\bibfield  {journal} {\bibinfo  {journal}
  {Journal of Magnetism and Magnetic Materials}\ }\textbf {\bibinfo {volume}
  {138}},\ \bibinfo {pages} {255} (\bibinfo {year} {1994})}\BibitemShut
  {NoStop}%
\bibitem [{\citenamefont {McGrouther}\ \emph {et~al.}(2016)\citenamefont
  {McGrouther}, \citenamefont {Lamb}, \citenamefont {Krajnak}, \citenamefont
  {McFadzean}, \citenamefont {McVitie}, \citenamefont {Stamps}, \citenamefont
  {Leonov}, \citenamefont {Bogdanov},\ and\ \citenamefont
  {Togawa}}]{mcgrouther2016internal}%
  \BibitemOpen
  \bibfield  {author} {\bibinfo {author} {\bibfnamefont {D.}~\bibnamefont
  {McGrouther}}, \bibinfo {author} {\bibfnamefont {R.}~\bibnamefont {Lamb}},
  \bibinfo {author} {\bibfnamefont {M.}~\bibnamefont {Krajnak}}, \bibinfo
  {author} {\bibfnamefont {S.}~\bibnamefont {McFadzean}}, \bibinfo {author}
  {\bibfnamefont {S.}~\bibnamefont {McVitie}}, \bibinfo {author} {\bibfnamefont
  {R.}~\bibnamefont {Stamps}}, \bibinfo {author} {\bibfnamefont
  {A.}~\bibnamefont {Leonov}}, \bibinfo {author} {\bibfnamefont
  {A.}~\bibnamefont {Bogdanov}}, \ and\ \bibinfo {author} {\bibfnamefont
  {Y.}~\bibnamefont {Togawa}},\ }\href@noop {} {\bibfield  {journal} {\bibinfo
  {journal} {New Journal of Physics}\ }\textbf {\bibinfo {volume} {18}},\
  \bibinfo {pages} {095004} (\bibinfo {year} {2016})}\BibitemShut {NoStop}%
\bibitem [{\citenamefont {Birch}\ \emph {et~al.}(2020)\citenamefont {Birch},
  \citenamefont {Cort{\'e}s-Ortu{\~n}o}, \citenamefont {Turnbull},
  \citenamefont {Wilson}, \citenamefont {Gro{\ss}}, \citenamefont {Tr{\"a}ger},
  \citenamefont {Laurenson}, \citenamefont {Bukin}, \citenamefont {Moody},
  \citenamefont {Weigand} \emph {et~al.}}]{birch2020real}%
  \BibitemOpen
  \bibfield  {author} {\bibinfo {author} {\bibfnamefont {M.}~\bibnamefont
  {Birch}}, \bibinfo {author} {\bibfnamefont {D.}~\bibnamefont
  {Cort{\'e}s-Ortu{\~n}o}}, \bibinfo {author} {\bibfnamefont {L.}~\bibnamefont
  {Turnbull}}, \bibinfo {author} {\bibfnamefont {M.}~\bibnamefont {Wilson}},
  \bibinfo {author} {\bibfnamefont {F.}~\bibnamefont {Gro{\ss}}}, \bibinfo
  {author} {\bibfnamefont {N.}~\bibnamefont {Tr{\"a}ger}}, \bibinfo {author}
  {\bibfnamefont {A.}~\bibnamefont {Laurenson}}, \bibinfo {author}
  {\bibfnamefont {N.}~\bibnamefont {Bukin}}, \bibinfo {author} {\bibfnamefont
  {S.}~\bibnamefont {Moody}}, \bibinfo {author} {\bibfnamefont
  {M.}~\bibnamefont {Weigand}},  \emph {et~al.},\ }\href@noop {} {\bibfield
  {journal} {\bibinfo  {journal} {Nature Communications}\ }\textbf {\bibinfo
  {volume} {11}},\ \bibinfo {pages} {1} (\bibinfo {year} {2020})}\BibitemShut
  {NoStop}%
\bibitem [{\citenamefont {Leonov}\ \emph {et~al.}(2016)\citenamefont {Leonov},
  \citenamefont {Togawa}, \citenamefont {Monchesky}, \citenamefont {Bogdanov},
  \citenamefont {Kishine}, \citenamefont {Kousaka}, \citenamefont {Miyagawa},
  \citenamefont {Koyama}, \citenamefont {Akimitsu}, \citenamefont {Koyama}
  \emph {et~al.}}]{leonov2016chiral}%
  \BibitemOpen
  \bibfield  {author} {\bibinfo {author} {\bibfnamefont {A.}~\bibnamefont
  {Leonov}}, \bibinfo {author} {\bibfnamefont {Y.}~\bibnamefont {Togawa}},
  \bibinfo {author} {\bibfnamefont {T.}~\bibnamefont {Monchesky}}, \bibinfo
  {author} {\bibfnamefont {A.}~\bibnamefont {Bogdanov}}, \bibinfo {author}
  {\bibfnamefont {J.-i.}\ \bibnamefont {Kishine}}, \bibinfo {author}
  {\bibfnamefont {Y.}~\bibnamefont {Kousaka}}, \bibinfo {author} {\bibfnamefont
  {M.}~\bibnamefont {Miyagawa}}, \bibinfo {author} {\bibfnamefont
  {T.}~\bibnamefont {Koyama}}, \bibinfo {author} {\bibfnamefont
  {J.}~\bibnamefont {Akimitsu}}, \bibinfo {author} {\bibfnamefont
  {T.}~\bibnamefont {Koyama}},  \emph {et~al.},\ }\href@noop {} {\bibfield
  {journal} {\bibinfo  {journal} {Physical Review Letters}\ }\textbf {\bibinfo
  {volume} {117}},\ \bibinfo {pages} {087202} (\bibinfo {year}
  {2016})}\BibitemShut {NoStop}%
\bibitem [{\citenamefont {Vansteenkiste}\ \emph {et~al.}(2014)\citenamefont
  {Vansteenkiste}, \citenamefont {Leliaert}, \citenamefont {Dvornik},
  \citenamefont {Helsen}, \citenamefont {Garcia-Sanchez},\ and\ \citenamefont
  {Van~Waeyenberge}}]{vansteenkiste2014design}%
  \BibitemOpen
  \bibfield  {author} {\bibinfo {author} {\bibfnamefont {A.}~\bibnamefont
  {Vansteenkiste}}, \bibinfo {author} {\bibfnamefont {J.}~\bibnamefont
  {Leliaert}}, \bibinfo {author} {\bibfnamefont {M.}~\bibnamefont {Dvornik}},
  \bibinfo {author} {\bibfnamefont {M.}~\bibnamefont {Helsen}}, \bibinfo
  {author} {\bibfnamefont {F.}~\bibnamefont {Garcia-Sanchez}}, \ and\ \bibinfo
  {author} {\bibfnamefont {B.}~\bibnamefont {Van~Waeyenberge}},\ }\href@noop {}
  {\bibfield  {journal} {\bibinfo  {journal} {AIP Advances}\ }\textbf {\bibinfo
  {volume} {4}},\ \bibinfo {pages} {107133} (\bibinfo {year}
  {2014})}\BibitemShut {NoStop}%
\bibitem [{\citenamefont {Takagi}\ \emph {et~al.}(2017)\citenamefont {Takagi},
  \citenamefont {Morikawa}, \citenamefont {Karube}, \citenamefont {Kanazawa},
  \citenamefont {Shibata}, \citenamefont {Tatara}, \citenamefont {Tokunaga},
  \citenamefont {Arima}, \citenamefont {Taguchi}, \citenamefont {Tokura} \emph
  {et~al.}}]{takagi2017spin}%
  \BibitemOpen
  \bibfield  {author} {\bibinfo {author} {\bibfnamefont {R.}~\bibnamefont
  {Takagi}}, \bibinfo {author} {\bibfnamefont {D.}~\bibnamefont {Morikawa}},
  \bibinfo {author} {\bibfnamefont {K.}~\bibnamefont {Karube}}, \bibinfo
  {author} {\bibfnamefont {N.}~\bibnamefont {Kanazawa}}, \bibinfo {author}
  {\bibfnamefont {K.}~\bibnamefont {Shibata}}, \bibinfo {author} {\bibfnamefont
  {G.}~\bibnamefont {Tatara}}, \bibinfo {author} {\bibfnamefont
  {Y.}~\bibnamefont {Tokunaga}}, \bibinfo {author} {\bibfnamefont
  {T.}~\bibnamefont {Arima}}, \bibinfo {author} {\bibfnamefont
  {Y.}~\bibnamefont {Taguchi}}, \bibinfo {author} {\bibfnamefont
  {Y.}~\bibnamefont {Tokura}},  \emph {et~al.},\ }\href@noop {} {\bibfield
  {journal} {\bibinfo  {journal} {Physical Review B}\ }\textbf {\bibinfo
  {volume} {95}},\ \bibinfo {pages} {220406} (\bibinfo {year}
  {2017})}\BibitemShut {NoStop}%
\bibitem [{\citenamefont {M{\"u}ller}\ \emph {et~al.}(2017)\citenamefont
  {M{\"u}ller}, \citenamefont {Rajeswari}, \citenamefont {Huang}, \citenamefont
  {Murooka}, \citenamefont {R{\o}nnow}, \citenamefont {Carbone},\ and\
  \citenamefont {Rosch}}]{muller2017magnetic}%
  \BibitemOpen
  \bibfield  {author} {\bibinfo {author} {\bibfnamefont {J.}~\bibnamefont
  {M{\"u}ller}}, \bibinfo {author} {\bibfnamefont {J.}~\bibnamefont
  {Rajeswari}}, \bibinfo {author} {\bibfnamefont {P.}~\bibnamefont {Huang}},
  \bibinfo {author} {\bibfnamefont {Y.}~\bibnamefont {Murooka}}, \bibinfo
  {author} {\bibfnamefont {H.~M.}\ \bibnamefont {R{\o}nnow}}, \bibinfo {author}
  {\bibfnamefont {F.}~\bibnamefont {Carbone}}, \ and\ \bibinfo {author}
  {\bibfnamefont {A.}~\bibnamefont {Rosch}},\ }\href@noop {} {\bibfield
  {journal} {\bibinfo  {journal} {Physical Review Letters}\ }\textbf {\bibinfo
  {volume} {119}},\ \bibinfo {pages} {137201} (\bibinfo {year}
  {2017})}\BibitemShut {NoStop}%
\bibitem [{\citenamefont {Loudon}\ \emph {et~al.}(2018)\citenamefont {Loudon},
  \citenamefont {Leonov}, \citenamefont {Bogdanov}, \citenamefont {Hatnean},\
  and\ \citenamefont {Balakrishnan}}]{loudon2017direct}%
  \BibitemOpen
  \bibfield  {author} {\bibinfo {author} {\bibfnamefont {J.~C.}\ \bibnamefont
  {Loudon}}, \bibinfo {author} {\bibfnamefont {A.}~\bibnamefont {Leonov}},
  \bibinfo {author} {\bibfnamefont {A.}~\bibnamefont {Bogdanov}}, \bibinfo
  {author} {\bibfnamefont {M.~C.}\ \bibnamefont {Hatnean}}, \ and\ \bibinfo
  {author} {\bibfnamefont {G.}~\bibnamefont {Balakrishnan}},\ }\href@noop {}
  {\bibfield  {journal} {\bibinfo  {journal} {Physical Review B}\ }\textbf
  {\bibinfo {volume} {97}},\ \bibinfo {pages} {134403} (\bibinfo {year}
  {2018})}\BibitemShut {NoStop}%
\bibitem [{\citenamefont {Rybakov}\ \emph {et~al.}(2016)\citenamefont
  {Rybakov}, \citenamefont {Borisov}, \citenamefont {Bl{\"u}gel},\ and\
  \citenamefont {Kiselev}}]{rybakov2016new}%
  \BibitemOpen
  \bibfield  {author} {\bibinfo {author} {\bibfnamefont {F.~N.}\ \bibnamefont
  {Rybakov}}, \bibinfo {author} {\bibfnamefont {A.~B.}\ \bibnamefont
  {Borisov}}, \bibinfo {author} {\bibfnamefont {S.}~\bibnamefont {Bl{\"u}gel}},
  \ and\ \bibinfo {author} {\bibfnamefont {N.~S.}\ \bibnamefont {Kiselev}},\
  }\href@noop {} {\bibfield  {journal} {\bibinfo  {journal} {New Journal of
  Physics}\ }\textbf {\bibinfo {volume} {18}},\ \bibinfo {pages} {045002}
  (\bibinfo {year} {2016})}\BibitemShut {NoStop}%
\bibitem [{\citenamefont {Rybakov}\ \emph {et~al.}(2015)\citenamefont
  {Rybakov}, \citenamefont {Borisov}, \citenamefont {Bl{\"u}gel},\ and\
  \citenamefont {Kiselev}}]{rybakov2015new}%
  \BibitemOpen
  \bibfield  {author} {\bibinfo {author} {\bibfnamefont {F.~N.}\ \bibnamefont
  {Rybakov}}, \bibinfo {author} {\bibfnamefont {A.~B.}\ \bibnamefont
  {Borisov}}, \bibinfo {author} {\bibfnamefont {S.}~\bibnamefont {Bl{\"u}gel}},
  \ and\ \bibinfo {author} {\bibfnamefont {N.~S.}\ \bibnamefont {Kiselev}},\
  }\href@noop {} {\bibfield  {journal} {\bibinfo  {journal} {Physical Review
  Letters}\ }\textbf {\bibinfo {volume} {115}},\ \bibinfo {pages} {117201}
  (\bibinfo {year} {2015})}\BibitemShut {NoStop}%
\bibitem [{\citenamefont {Zheng}\ \emph {et~al.}(2018)\citenamefont {Zheng},
  \citenamefont {Rybakov}, \citenamefont {Borisov}, \citenamefont {Song},
  \citenamefont {Wang}, \citenamefont {Li}, \citenamefont {Du}, \citenamefont
  {Kiselev}, \citenamefont {Caron}, \citenamefont {Kov{\'a}cs} \emph
  {et~al.}}]{zheng2018experimental}%
  \BibitemOpen
  \bibfield  {author} {\bibinfo {author} {\bibfnamefont {F.}~\bibnamefont
  {Zheng}}, \bibinfo {author} {\bibfnamefont {F.~N.}\ \bibnamefont {Rybakov}},
  \bibinfo {author} {\bibfnamefont {A.~B.}\ \bibnamefont {Borisov}}, \bibinfo
  {author} {\bibfnamefont {D.}~\bibnamefont {Song}}, \bibinfo {author}
  {\bibfnamefont {S.}~\bibnamefont {Wang}}, \bibinfo {author} {\bibfnamefont
  {Z.-A.}\ \bibnamefont {Li}}, \bibinfo {author} {\bibfnamefont
  {H.}~\bibnamefont {Du}}, \bibinfo {author} {\bibfnamefont {N.~S.}\
  \bibnamefont {Kiselev}}, \bibinfo {author} {\bibfnamefont {J.}~\bibnamefont
  {Caron}}, \bibinfo {author} {\bibfnamefont {A.}~\bibnamefont {Kov{\'a}cs}},
  \emph {et~al.},\ }\href@noop {} {\bibfield  {journal} {\bibinfo  {journal}
  {Nature Nanotechnology}\ ,\ \bibinfo {pages} {1}} (\bibinfo {year}
  {2018})}\BibitemShut {NoStop}%
\bibitem [{\citenamefont {Ahmed}\ \emph {et~al.}(2018)\citenamefont {Ahmed},
  \citenamefont {Rowland}, \citenamefont {Esser}, \citenamefont {Dunsiger},
  \citenamefont {McComb}, \citenamefont {Randeria},\ and\ \citenamefont
  {Kawakami}}]{ahmed2018chiral}%
  \BibitemOpen
  \bibfield  {author} {\bibinfo {author} {\bibfnamefont {A.~S.}\ \bibnamefont
  {Ahmed}}, \bibinfo {author} {\bibfnamefont {J.}~\bibnamefont {Rowland}},
  \bibinfo {author} {\bibfnamefont {B.~D.}\ \bibnamefont {Esser}}, \bibinfo
  {author} {\bibfnamefont {S.~R.}\ \bibnamefont {Dunsiger}}, \bibinfo {author}
  {\bibfnamefont {D.~W.}\ \bibnamefont {McComb}}, \bibinfo {author}
  {\bibfnamefont {M.}~\bibnamefont {Randeria}}, \ and\ \bibinfo {author}
  {\bibfnamefont {R.~K.}\ \bibnamefont {Kawakami}},\ }\href@noop {} {\bibfield
  {journal} {\bibinfo  {journal} {Physical Review Materials}\ }\textbf
  {\bibinfo {volume} {2}},\ \bibinfo {pages} {041401} (\bibinfo {year}
  {2018})}\BibitemShut {NoStop}%
\bibitem [{\citenamefont {M{\"u}ller}\ \emph {et~al.}(2020)\citenamefont
  {M{\"u}ller}, \citenamefont {Rybakov}, \citenamefont {J{\'o}nsson},
  \citenamefont {Bl{\"u}gel},\ and\ \citenamefont
  {Kiselev}}]{muller2020coupled}%
  \BibitemOpen
  \bibfield  {author} {\bibinfo {author} {\bibfnamefont {G.~P.}\ \bibnamefont
  {M{\"u}ller}}, \bibinfo {author} {\bibfnamefont {F.~N.}\ \bibnamefont
  {Rybakov}}, \bibinfo {author} {\bibfnamefont {H.}~\bibnamefont
  {J{\'o}nsson}}, \bibinfo {author} {\bibfnamefont {S.}~\bibnamefont
  {Bl{\"u}gel}}, \ and\ \bibinfo {author} {\bibfnamefont {N.~S.}\ \bibnamefont
  {Kiselev}},\ }\href@noop {} {\bibfield  {journal} {\bibinfo  {journal}
  {Physical Review B}\ }\textbf {\bibinfo {volume} {101}},\ \bibinfo {pages}
  {184405} (\bibinfo {year} {2020})}\BibitemShut {NoStop}%
\bibitem [{\citenamefont {Leonov}\ \emph {et~al.}(2018)\citenamefont {Leonov},
  \citenamefont {Bogdanov},\ and\ \citenamefont {Inoue}}]{leonov2018toggle}%
  \BibitemOpen
  \bibfield  {author} {\bibinfo {author} {\bibfnamefont {A.~O.}\ \bibnamefont
  {Leonov}}, \bibinfo {author} {\bibfnamefont {A.~N.}\ \bibnamefont
  {Bogdanov}}, \ and\ \bibinfo {author} {\bibfnamefont {K.}~\bibnamefont
  {Inoue}},\ }\href@noop {} {\bibfield  {journal} {\bibinfo  {journal}
  {Physical Review B}\ }\textbf {\bibinfo {volume} {98}},\ \bibinfo {pages}
  {060411} (\bibinfo {year} {2018})}\BibitemShut {NoStop}%
\bibitem [{\citenamefont {Vlasov}\ \emph {et~al.}(2020)\citenamefont {Vlasov},
  \citenamefont {Uzdin},\ and\ \citenamefont {Leonov}}]{vlasov2020skyrmion}%
  \BibitemOpen
  \bibfield  {author} {\bibinfo {author} {\bibfnamefont {S.~M.}\ \bibnamefont
  {Vlasov}}, \bibinfo {author} {\bibfnamefont {V.~M.}\ \bibnamefont {Uzdin}}, \
  and\ \bibinfo {author} {\bibfnamefont {A.~O.}\ \bibnamefont {Leonov}},\
  }\href@noop {} {\bibfield  {journal} {\bibinfo  {journal} {Journal of
  Physics: Condensed Matter}\ }\textbf {\bibinfo {volume} {32}},\ \bibinfo
  {pages} {185801} (\bibinfo {year} {2020})}\BibitemShut {NoStop}%
\bibitem [{\citenamefont {Ukleev}\ \emph {et~al.}(2019)\citenamefont {Ukleev},
  \citenamefont {Yamasaki}, \citenamefont {Morikawa}, \citenamefont {Karube},
  \citenamefont {Shibata}, \citenamefont {Tokunaga}, \citenamefont {Okamura},
  \citenamefont {Amemiya}, \citenamefont {Valvidares}, \citenamefont {Nakao}
  \emph {et~al.}}]{ukleev2019element}%
  \BibitemOpen
  \bibfield  {author} {\bibinfo {author} {\bibfnamefont {V.}~\bibnamefont
  {Ukleev}}, \bibinfo {author} {\bibfnamefont {Y.}~\bibnamefont {Yamasaki}},
  \bibinfo {author} {\bibfnamefont {D.}~\bibnamefont {Morikawa}}, \bibinfo
  {author} {\bibfnamefont {K.}~\bibnamefont {Karube}}, \bibinfo {author}
  {\bibfnamefont {K.}~\bibnamefont {Shibata}}, \bibinfo {author} {\bibfnamefont
  {Y.}~\bibnamefont {Tokunaga}}, \bibinfo {author} {\bibfnamefont
  {Y.}~\bibnamefont {Okamura}}, \bibinfo {author} {\bibfnamefont
  {K.}~\bibnamefont {Amemiya}}, \bibinfo {author} {\bibfnamefont
  {M.}~\bibnamefont {Valvidares}}, \bibinfo {author} {\bibfnamefont
  {H.}~\bibnamefont {Nakao}},  \emph {et~al.},\ }\href@noop {} {\bibfield
  {journal} {\bibinfo  {journal} {Physical Review B}\ }\textbf {\bibinfo
  {volume} {99}},\ \bibinfo {pages} {144408} (\bibinfo {year}
  {2019})}\BibitemShut {NoStop}%
\end{thebibliography}
\end{document}